\documentclass[fleqn,usenatbib]{mnras}

\usepackage{newtxtext,newtxmath}
\usepackage{xcolor}
\usepackage{footnote}

\usepackage[T1]{fontenc}

\DeclareRobustCommand{\VAN}[3]{#2}
\let\VANthebibliography\thebibliography
\def\thebibliography{\DeclareRobustCommand{\VAN}[3]{##3}\VANthebibliography}

\usepackage{graphicx}	
\usepackage{amsmath}	

\title[Red Quasars in X-Rays]{Accretion and Obscuration in Merger-Dominated Luminous Red Quasars}

\author[E. Glikman et al.]{
Eilat Glikman,$^{1}$\thanks{E-mail: eglikman@middlebury.edu}
Stephanie LaMassa,$^{2}$
Enrico Piconcelli,$^{3}$
Luca Zappacosta$^{3}$
and Mark Lacy$^{4}$
\\
$^{1}$Department of Physics, Middlebury College, Middlebury, VT 05753, USA\\
$^{2}$Space Telescope Science Institute, 3700 San Martin Drive, Baltimore MD, 21218, USA\\
$^{3}$Osservatorio Astronomico di Roma (INAF), via Frascati 33, 00040 Monte Porzio Catone (Roma), Italy\\
$^{4}$National Radio Astronomy Observatory, Charlottesville, VA, USA
}

\date{Accepted XXX. Received YYY; in original form ZZZ}

\pubyear{2024}

\begin{document}
\label{firstpage}
\pagerange{\pageref{firstpage}--\pageref{lastpage}}
\maketitle

\begin{abstract}
We present an analysis of the X-ray properties 10 luminous, dust-reddened quasars from the FIRST-2MASS (F2M) survey based on new and archival {\em Chandra} observations. 
These systems are interpreted to be young, transitional objects predicted by merger-driven models of quasar/galaxy co-evolution. 
The sources have been well-studied from the optical through mid-infrared, have Eddington ratios above 0.1, and possess high-resolution imaging, most of which shows disturbed morphologies indicative of a recent or ongoing merger. 
When combined with previous X-ray studies of five other F2M red quasars, we find that the sources, especially those hosted by mergers, have moderate to high column densities ($N_H \simeq 10^{22.5-23.5}$ cm$^{-2}$) and Eddington ratios high enough to enable radiation pressure to blow out the obscuring material. 
We confirm previous findings that red quasars have dust-to-gas ratios that are significantly lower than the value for the Milky Way's interstellar medium, especially when hosted by a merger.  
The dust-to-gas ratio for two red quasars that lack evidence for merging morphology is consistent with the Milky Way and they do not meet the radiative feedback conditions for blowout. 
These findings support the picture of quasar/galaxy co-evolution in which a merger results in feeding of and feedback from an AGN.
We compare the F2M red quasars to other obscured and reddened quasar populations in the literature, finding that, although morphological information is lacking, nearly all such samples meet blowout conditions and exhibit outflow signatures suggestive of winds and feedback. 
\end{abstract}
 



\section{Introduction} \label{sec:intro}

A complete picture of galaxy evolution must include the growth of supermassive black holes (SMBHs) at their centres, as evidence suggests a formation and evolutionary relationship between the two. 
The ubiquity of SMBHs in the centres of galaxies \citep{Faber97}, the tight $M_{BH} - \sigma$ relation \citep{Gebhardt00,Ferrarese00}, and the contemporaneous peak in star formation and black hole growth over cosmic history \citep{HopkinsBeacom06} all point to an energy exchange, or “feedback”, between the black holes and their hosts. 
This feedback from active galactic nuclei (AGN) is still poorly understood, and may come in the form of radiation, winds, outflows, and/or jets \citep{Fabian12}.

One way to explain these observations is through major galaxy mergers that induce both SMBH accretion and circumnuclear star-formation, resulting in large amounts of dust and gas that obscure much of the SMBH’s growth \citep{Sanders88,Hopkins06}. 
According to this model, the obscuring dust is eventually cleared by powerful quasar winds, revealing luminous, unreddened emission from the quasar. 
In this scenario dust-reddened (or “red”) quasars represent a crucial early phase in SMBH/galaxy co-evolution: the transition from a dust-enshrouded core to a typical, unobscured quasar. 
In the context of this picture, the reddened phase represents a key component of SMBH growth with the potential to reveal the physics of feedback once the quasar becomes luminous enough to blow away the circumnuclear material. 

Recently, samples of heavily reddened quasars have been shown to fit into this scenario as the long-sought transitioning population \citep[e.g.,][]{Banerji12,Tsai15,LaMassa17}. 
A red quasar sample constructed from the cross-matching of the Faint Images of the Radio Sky at Twenty cm \citep[FIRST;][]{Becker95} survey to the Two-Micron All-Sky Survey \citep[2MASS;][]{Skrutskie06}, applying red optical-to-near infrared colour cuts, and spectroscopically confirming broad-line (Type 1) sources yielded $\sim$130 objects that span a broad range of redshifts $(0.1 < z < 3)$ and reddenings \citep[$0.1 < E(B - V)  < 1.5$;][hereafter called F2M red quasars]{Glikman04,Glikman07,Urrutia09,Glikman12,Glikman13}.
Extensive observations of F2M red quasars show that they are in a transitional phase of a merger-driven process: {\em Hubble} Space Telescope (HST) images show mergers are very common \citep[$>80\%$][]{Urrutia08,Glikman15}; they have high accretion rates \citep[$L/L_{\rm Edd} \gtrsim 0.3$][]{Kim15}; their BH masses are under-massive compared to their hosts, suggesting they have not finished growing \citep{Urrutia12}; and a high fraction of them exhibit outflows and winds via the presence of blue-shifted broad absorption lines in low-ionization species \citep[i.e., LoBALS and FeLoBALs make up $>60\%$ of F2M red quasars compared to 5\% in the general quasar population;][]{Urrutia09} indicative of winds and outflows.  
More recently, integral field spectroscopy of three F2M red quasars show bi-conal superbubbles in [\ion{O}{iii}] emission, catching the short-lived ``break-out'' phase \citep{Shen23}. 

One way to determine whether an AGN is in the radiatively-driven ``blow-out'' phase is by comparing its Eddington ratio (${\lambda }_{\mathrm{Edd}}=L/{L}_{\mathrm{Edd}}$) to the hydrogen column density ($N_H$). 
A study of hard X-ray-selected local ($z<0.05$) AGN showed that they are either completely obscured, with $\mathrm{log}(\lambda_{\rm Edd} )\lesssim -1.5$ and $N_H > 10^{22}$ cm$^{-2}$, or largely unobscured, with $\mathrm{log}(\lambda_{\rm Edd} )\gtrsim -1.5$ and $N_H < 10^{22}$ cm$^{-2}$ \citep{Ricci17}. 
There exist a unique set of conditions whereby an AGN has sufficiently high $\lambda_{\rm Edd}$ and a not-too-high $N_H$ to blow out the dust and gas \citep{Fabian08,Ishibashi18}. 
Recently, \citet{Stacey22} used ALMA observations to show that reddened quasars with $E(B-V)>0.5$ reside in this ``blow-out'' region of $\lambda_{\rm Edd}$ vs.~$N_H$ space.

While we have measured black hole masses, Eddington ratios, and reddenings ($E(B-V)$) for all the F2M red quasars, our understanding of their X-ray properties has been deficient. 
Twelve F2M quasars were observed with {\em Chandra} in 2004 with $5 - 10$ ksec exposures \citep{Urrutia05}. 
While all of the sources show absorbed X-ray spectra, the detections were mostly too low-count (all but one had $<100$ counts) for detailed spectral analysis and the ancillary data for F2M red quasars had not been obtained making it difficult to draw conclusions. 

More recently, we obtained high-quality X-ray spectra of four F2M red quasars with {\em XMM-Newton} and {\em NuSTAR}, as well as archival {\em Chandra} data \citep[hereafter, L16 and G17, respectively]{LaMassa16b,Glikman17a}; three of these have {\em HST} images showing merging hosts.
We found that these three sources fall squarely in the blowout region of the the $\lambda_{\rm Edd}$ vs. $N_H$ diagram \citep{Glikman17b}. 
The source that lies outside of the blowout region lacks morphological information, and has a dust-to-gas ratio consistent with the Galactic value, possibly because it is obscured by dust lanes in its host galaxy. 
In addition, a fifth F2M red quasar (F2M~J0915) has a 3.2 ksec {\em Chandra} observation analyzed in \citet{Urrutia05} as well as high-spatial-resolution imaging revealing a merging host \citep{Urrutia08}.  
We list the properties of these F2M red quasars in Table \ref{tab:previous_sources} as a reference for the remainder of the paper. 

\defcitealias{LaMassa16b}{L16}
\defcitealias{Glikman17a}{G17}

\begin{table*}
\caption{Properties of Previously Studied F2M Red Quasars}
\label{tab:previous_sources}
\begin{tabular}{cccccccccc}
\hline
{Name}    & {R.A.}       & {Decl.}    & {Redshift} & {$E(B-V)$}    & {$\log L_{\rm bol}^\dagger$} & {$L/L_{\rm Edd}^\ddagger$} & {$\log{N_H}$} & {Merger?} & {Ref} \\ 
{}        & {(J2000)}    & {(J2000)}  & {}         & {(mag)}       & {(erg~s$^{-1}$)}     & {}                        &  {(cm$^{-2}$)} & {}      & {}    \\
\hline
F2M~J0830 & 08:30:11.12 & +37:59:51.8 & 0.414     &  $0.71\pm0.01$  &  $46.20\pm0.01$    &  $0.4\pm0.1$   & $22.32\pm0.04$       & Y & \citetalias{LaMassa16b} \\
F2M~J0915 & 09:15:01.70 & +24:18:12.2 & 0.842     &  $0.73\pm0.02$  & $47.696\pm0.006$   & $0.94\pm0.48$  & $22.8^{+0.2}_{-0.4}$ & Y & \citet{Urrutia05} \\
F2M~J1113 & 11:13:54.67 & +12:44:38.9 & 0.681     &  $1.26\pm0.01$  & $47.475\pm0.006$   & $2.29\pm0.42$  & $23.1\pm0.1$         & Y & \citetalias{Glikman17a} \\
F2M~J1227 & 12:27:49.15 & +32:14:59.0 & 0.137     & $0.828\pm0.003$ & $45.545\pm0.005$   &  $0.8\pm0.2$   & $21.5\pm0.1$         & ? & \citetalias{LaMassa16b} \\
F2M~J1656 & 16:56:47.11 & +38:21:36.7 & 0.732     & $0.519\pm0.004$ &  $46.81\pm0.01$    & $0.76\pm0.18$  & $23.0\pm0.1$         & Y & \citetalias{Glikman17a} \\
\hline
\multicolumn{9}{l}{$^\dagger$ Bolometric luminosities were determined by applying a bolometric correction of 7.6 \citep{Richards06} to the 6$\mu$m luminosity,}\\
\multicolumn{9}{l}{\phantom{a} which was determined by interpolating their WISE mid-infrared luminosities in the rest-frame.  }\\
\multicolumn{9}{l}{$^\ddagger$ As reported in \citet{Kim15} except for F2M~J0830 and F2M~J1227 which were obtained from \citetalias{Glikman17a}.}\\
\end{tabular}
\end{table*}

X-rays give the best measure of an AGN’s true underlying accretion luminosity, because they originate close to the black hole and penetrate gas and dust for all but the most obscured AGN ($N_H < 10^{24}$ cm$^{-2}$ and $E < 10$ keV).
Besides showing evidence for active radiative feedback, the studies in \citetalias{LaMassa16b} and \citetalias{Glikman17a} found that the three merger-hosted sources were best fit by an absorbed power-law model with a small fraction of the incident emission being leaked or scattered back into the line-of-sight (we refer to this as the `scattering fraction'; here, $f_{\rm scatt} = 1-7\%$) and moderate line-of-sight extinction ($N_H = 10^{22 - 23}$ cm$^{-2}$). 
Intriguingly, self-consistent physically-motivated model fitting with MYTorus \citep{Murphy09} exposes the presence of globally distributed gas suggesting a more complex environment than a simple absorber along the line-of-sight. 

In this paper we present X-ray observations for 10 additional F2M red quasars, which doubles the initial sample to more robustly verify the previous results. 
All the sources have high resolution imaging which enable us to tie host galaxy morphology to X-ray properties, including their potential for existing in a blowout phase. 
When optical magnitudes are discussed, we specify whether they are on the AB or Vega system via a subscript.
Uncertainties on X-ray parameters are reported as 90\% confidence limits. 
Throughout this work, we adopt the concordance $\Lambda$CDM cosmology with $H_0 = 70$~km~s$^{-1}$~Mpc$^{-1}$, $\Omega_M = 0.3$, and $\Omega_\Lambda = 0.7$. 

\section{The Sample and Observations} 

\subsection{Source Selection and Characteristics} \label{sec:sel} 

Of the $\sim130$ F2M red quasars, all have optical and/or near-infrared spectroscopy, as well as photometric coverage from ultraviolet to mid-infrared wavelengths and 27 have {\em HST}\footnote{24 red quasars have targeted {\em HST} imaging from \citet[13 objects]{Urrutia08} and \citet[11 objects]{Glikman15}, one red quasar was targeted in a snapshot {\em HST} program \citep{Marble03}, and another was serendipitously located in the background of another {\em HST} snapshot program (GO-11604).}
or other high resolution imaging (either targeted or serendipitous). 
For this study, we assembled a list of F2M red quasars that have the following observations in hand: (1) high resolution imaging from {\em HST} or other imaging; (2) optical and/or near-infrared spectra with at least one broad emission line enabling an estimate of a black-hole mass ($M_{BH}$). 
We further required that our targets yield at least 70 counts (see \S \ref{sec:chandra}) in $< 20$ ksec {\em Chandra} observation and identified eight sources that obeyed these criteria.
In addition, two sources were found in the background of archival {\em Chandra} observations, with one source appearing in two different datasets. 
Our sample, therefore consists of 10 F2M red quasars with new or archival {\em Chandra} observations.

\begin{figure*}
\begin{center}
\includegraphics[scale=0.35]{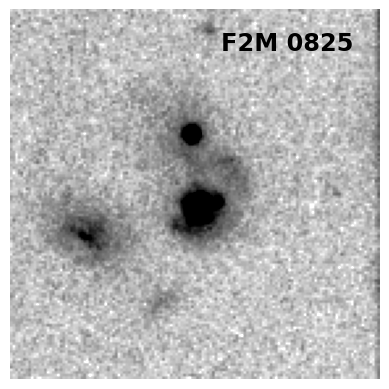} 
\includegraphics[scale=0.35]{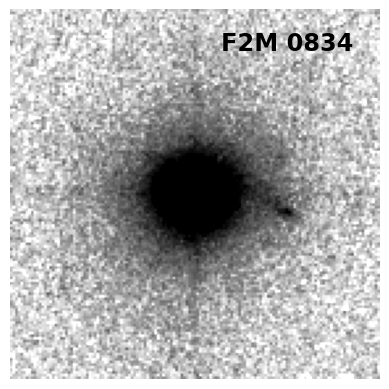}
\includegraphics[scale=0.35]{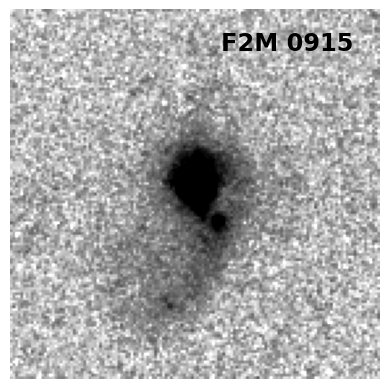} 
\includegraphics[scale=0.35]{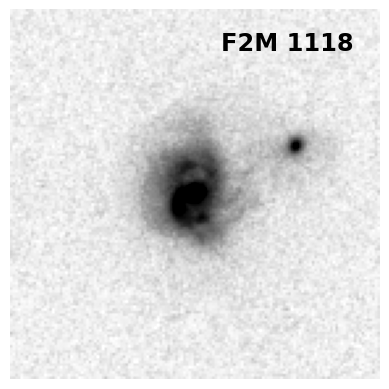}
\includegraphics[scale=0.35]{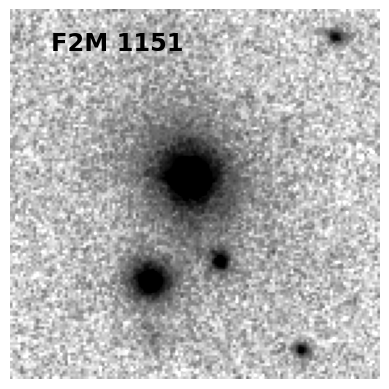}
\includegraphics[scale=0.35]{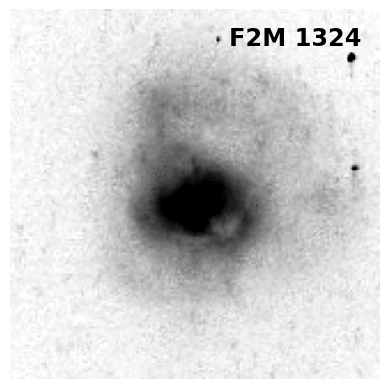}
\includegraphics[scale=0.35]{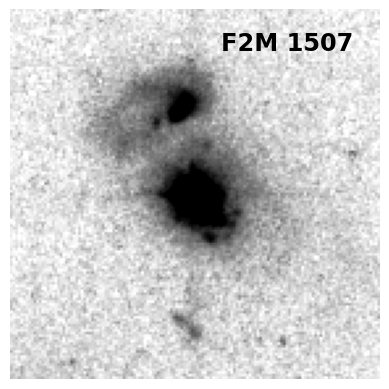}
\includegraphics[scale=0.35]{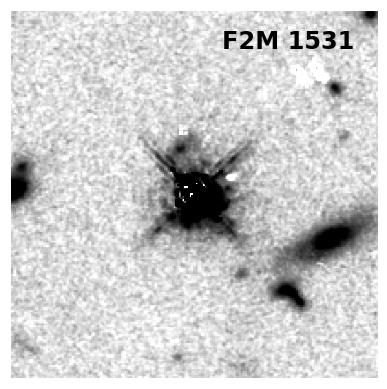}
\includegraphics[scale=0.35]{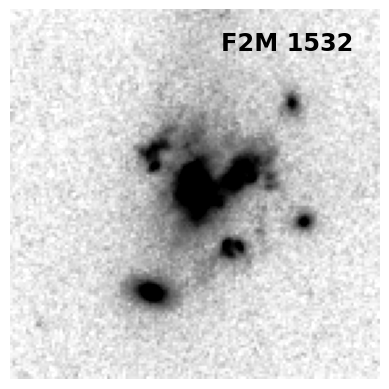}
\includegraphics[scale=0.35]{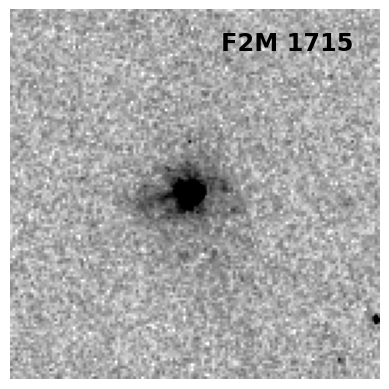}
\caption{
Image cutouts of 9 the 10 F2M red quasars presented in this work, excluding F2M~J1106, but including F2M~J0915 which was not presented in the analysis of \citetalias{LaMassa16b} or \citetalias{Glikman17a}. All data are from HST ACS camera with the F814W filter, except for F2M~J1531 which shows the PSF-subtracted image from \citet{Glikman15} from the WFC3/IR camera with the F160W filter. The source for each image is listed in the final column of Table \ref{tab:obs}. All images are $7\arcsec \times 7\arcsec$ except F2M~J1324, which is $8\arcsec \times 8\arcsec$ due to it having the lowest redshift ($z=0.205$) and larger angular size, and F2M~J1531, which is $8\arcsec \times 8\arcsec$.
\label{fig:images}}
\end{center} 
\end{figure*}

Figure \ref{fig:images} shows the {\em HST} image cutouts for these sources, as well as for F2M~J0915. The images were obtained from the archives, except for F2M~J1531 whose point spread function (PSF) subtracted WFC3/IR F160W image from \citet{Glikman15} is reproduced here. 
The morphology of F2M~J1106 is based on integral field spectroscopy (IFS) with the GMOS instrument showing bi-conal bubbles in the [\ion{O}{iii}] \citep[see][]{Shen23}.
Images of the remaining sources listed in Table \ref{tab:previous_sources} are shown in \citetalias{LaMassa16b} and \citetalias{Glikman17a}.

The fifth column of Tables \ref{tab:previous_sources} and \ref{tab:sources} lists the reddening of each quasar parametrized by the color excess, $E(B-V)$, which we determined by performing a linear fit in log space to the ratio of each red quasar spectrum, $f(\lambda)$, to an unreddened quasar template spectrum, i.e.  
\begin{equation}
\log{\left[ \frac{f(\lambda)}{f_0(\lambda)} \right]} = -\frac{k(\lambda) E(B-V)}{1.086}. 
\end{equation}
Here, $f_0(\lambda)$ is the optical-to-near-infrared quasar composite template from \citet{Glikman06} and $k(\lambda)$ is the Small Magellanic Cloud (SMC) dust extinction law from \citet{Gordon98}.
Although $E(B-V)$ values were already in-hand, we recomputed them for this work as newer spectra had been obtained for some sources which, in some cases, broadened the wavelength coverage or, in others, improved the signal-to-noise. The uncertainties on $E(B-V)$ were computed by heavily smoothing and perturbing the original spectrum by its own error array and re-fitting it to determine $E(B-V)$ 1000 times. The reported $E(B-V)$ uncertainty is then the standard deviation of that $E(B-V)$ distribution.

\begin{figure*}
\begin{center}
\includegraphics[scale=0.27]{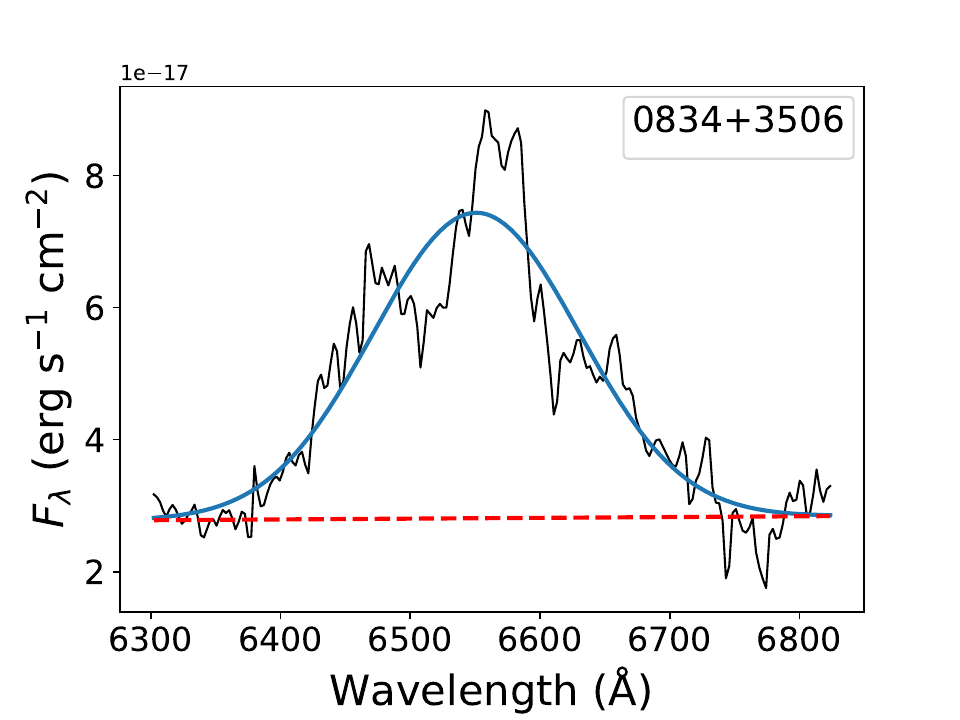}
\includegraphics[scale=0.27]{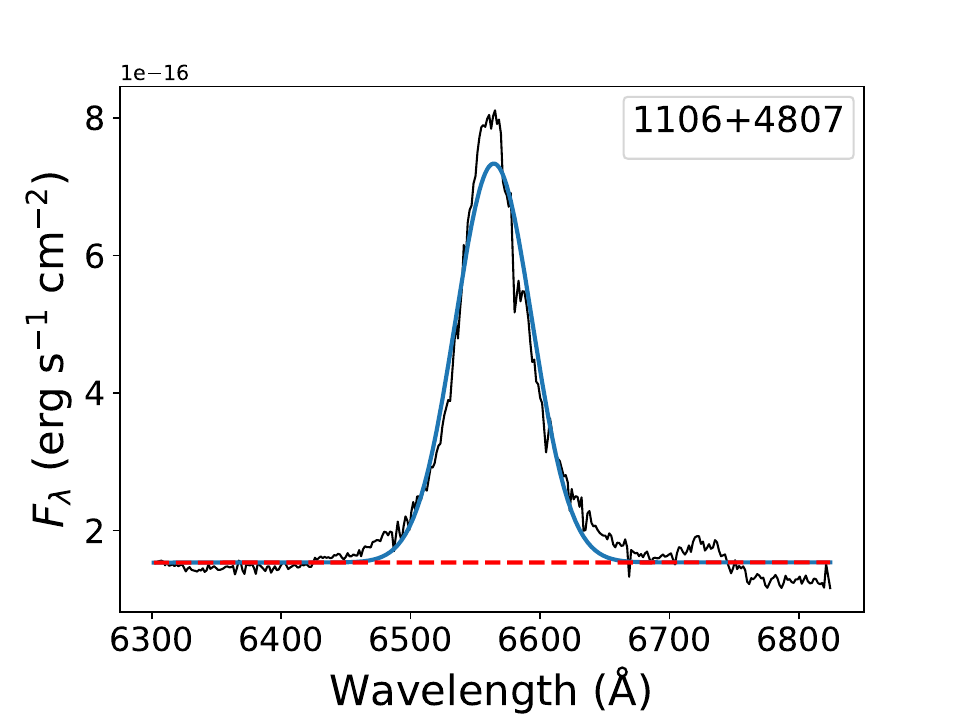} 
\includegraphics[scale=0.27]{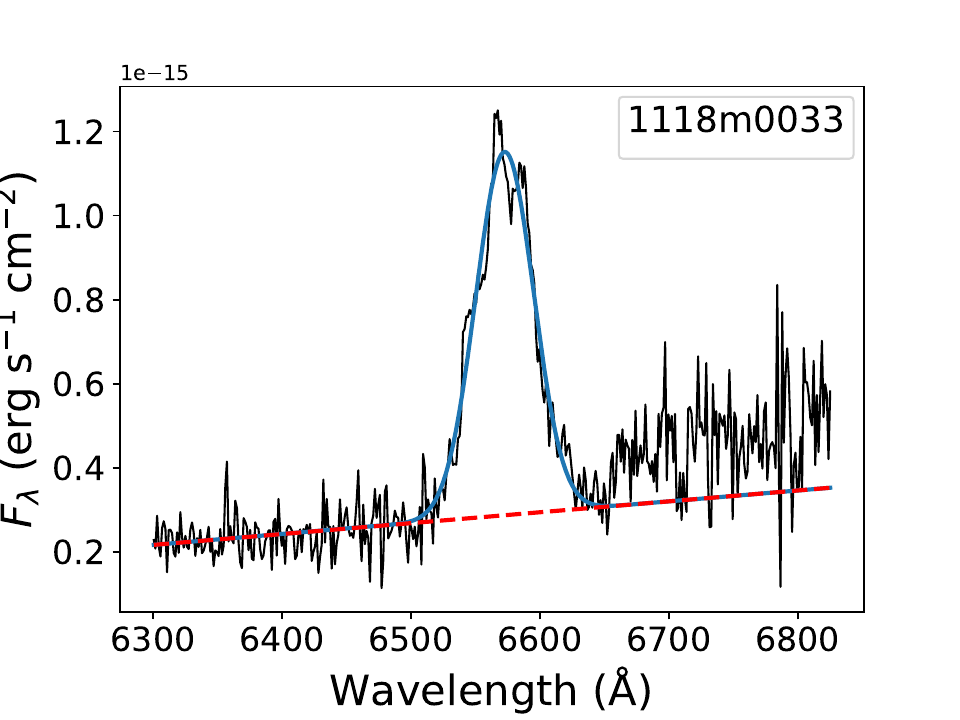}
\includegraphics[scale=0.27]{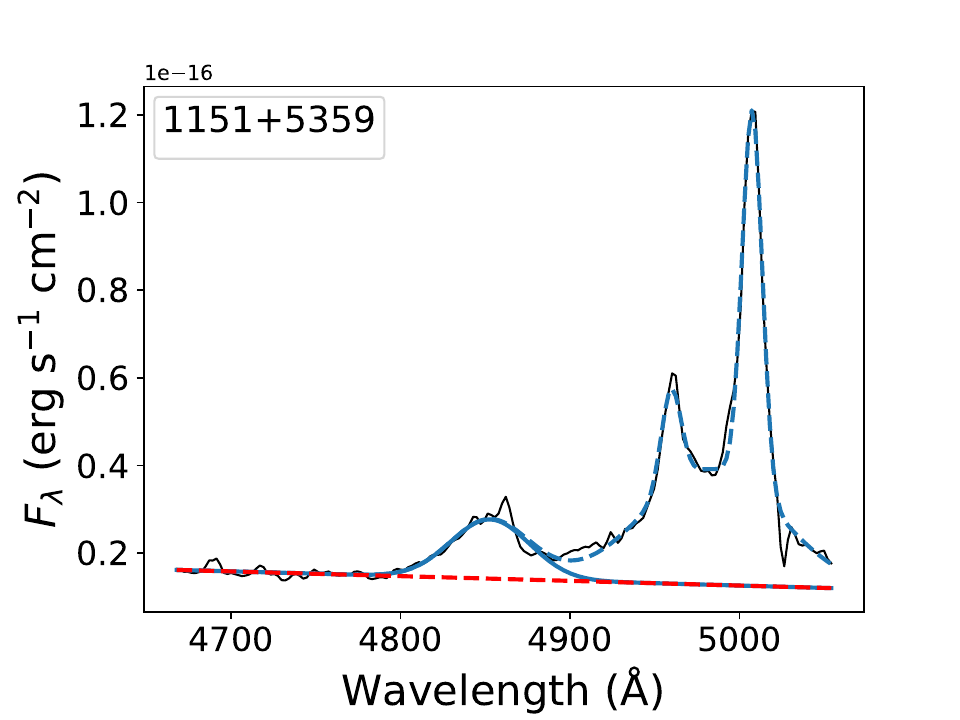}
\includegraphics[scale=0.27]{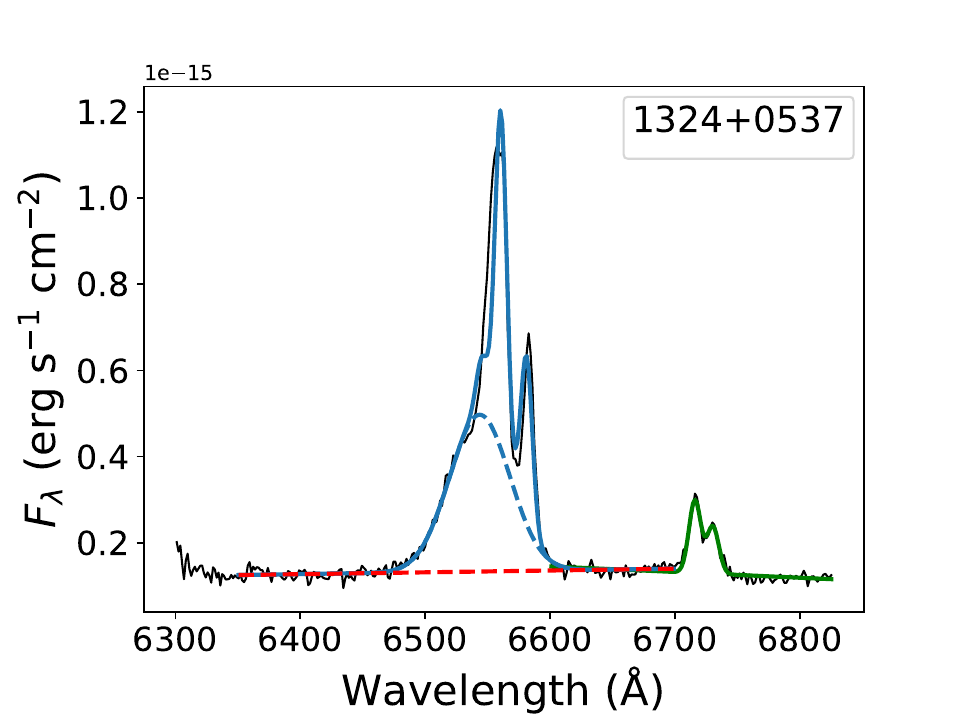}
\includegraphics[scale=0.27]{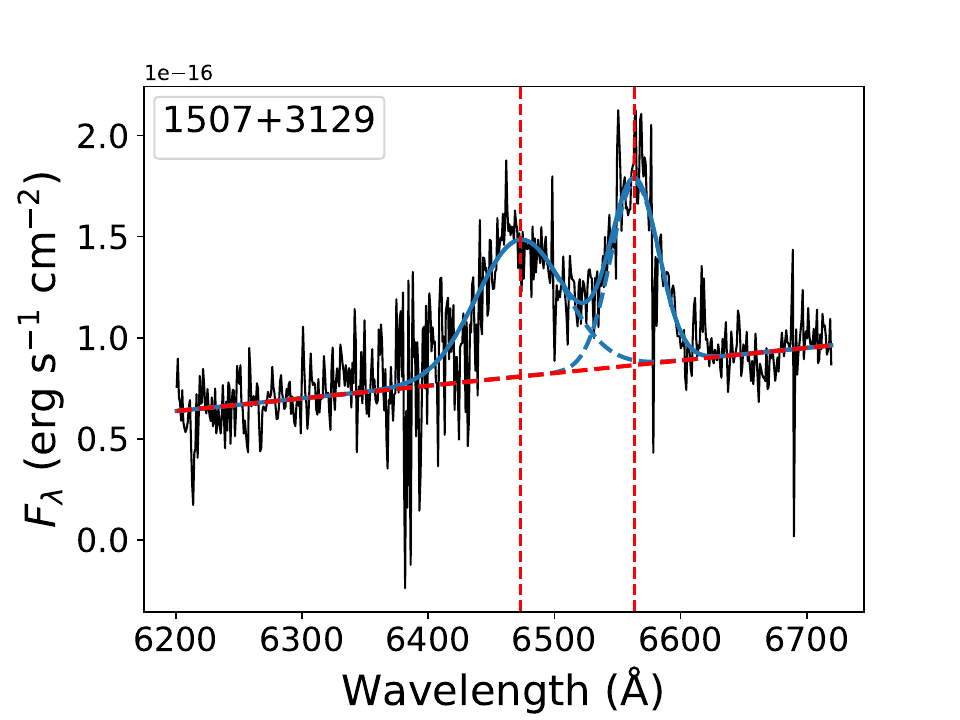}
\includegraphics[scale=0.27]{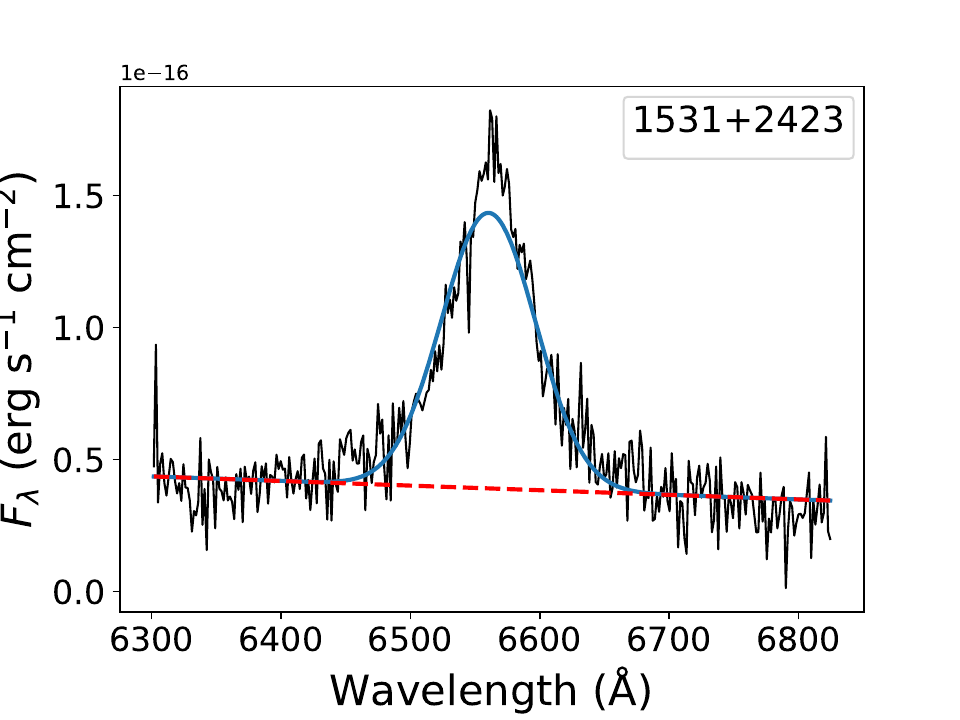}
\includegraphics[scale=0.27]{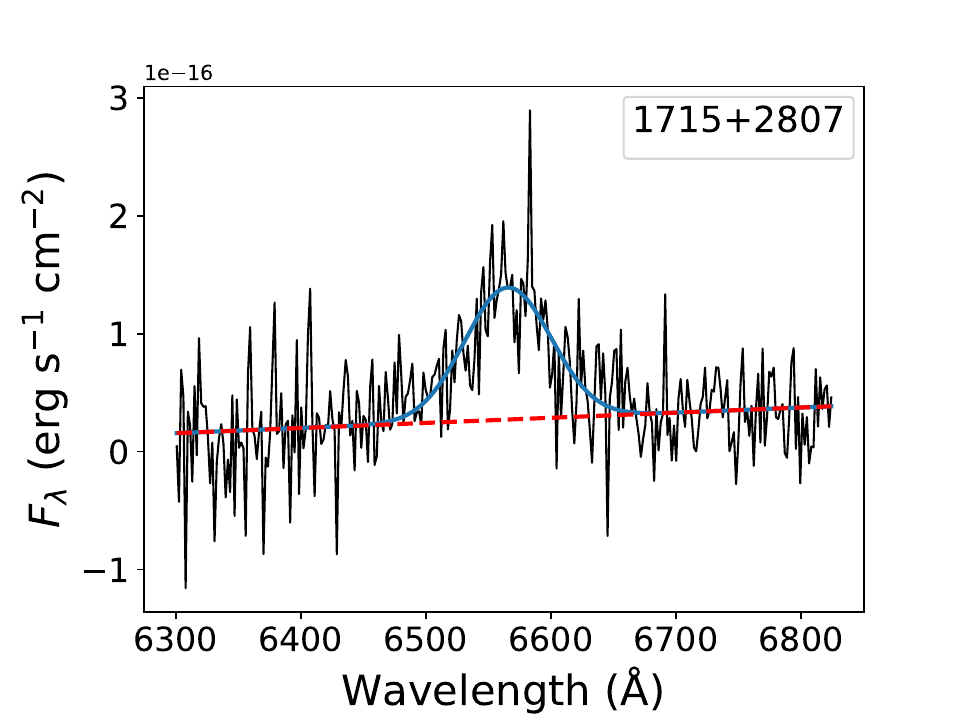}
\caption{
Gaussian fitting to emission lines for eight quasars in our sample that lack $M_{BH}$ from \citet{Kim15}. 
The black line shows the observed flux against rest wavelength.
The blue line is the best-fit model to the line profile. 
The sloped dotted red line is the continuum portion of the best-fit model. In most cases, a single Gaussian sufficiently fits the data. The three exceptions, from left to right, are: (1) F2M~J1151, whose [\ion{O}{iii}] line doublet is fit together with H$\beta$. (2) F2M~J1324, whose H$\alpha$ line is decomposed into a broad and narrow components and fit along with the [\ion{N}{ii}] nitrogen doublet. In this case the narrow line width is determined by fitting to the [\ion{S}{ii}] doublet shown in green. And, (3) F2M~J1507, whose broad H$\alpha$ line is double-peaked with a blue-shifted component separated by 91\AA\ (see Appendix \ref{appendix:f2m1507} for additional discussion of this source). 
\label{fig:mbh}}
\end{center}
\end{figure*}

We determine the black hole masses, $M_{BH}$, from a broad emission line in the quasars' spectra. 
Eight sources were analyzed using their broad emission line widths, either from H$\beta$ or H$\alpha$, using the line with the highest signal-to-noise ratio. 
We performed multi-component Gaussian fits to the lines, including narrow emission line components combined with a broad component. 
Figure \ref{fig:mbh} shows these line fits.

We use the established relations from \citet{Shen12}, 
\begin{equation}
\log \bigg(\frac{M_{\rm BH,vir}}{M_\odot} \bigg) = a + b \log \bigg(\frac{L_{5100}}{10^{44} \rm erg/s} \bigg) + c \log \bigg(\frac{v_{\rm FWHM}}{\rm km/s} \bigg),
\label{eqn:mbh}
\end{equation}
to compute $M_{BH}$ for each line species, employing the full-width at half maximum (FWHM) in km s$^{-1}$ for the velocity term. 
When the line used was H$\alpha$, we adopted the values $a=0.774$, $b=0.520$, $c=2.06$ for sources with $L_{5100}<10^{45.4}$ erg s$^{-1}$ and $a=1.390$, $b=0.555$, $c=1.873$ for for sources with $L_{5100}>10^{45.4}$ erg s$^{-1}$. 
For $M_{BH}$ estimates based on H$\beta$, we adopted the values $a=0.895$, $b=0.520$, $c=2.00$, which apply to sources with $L_{5100}<10^{45.4}$, erg s$^{-1}$ \citep[the $a$, $b$, $c$ coefficients are from the calibration of][]{Assef11}.  

Two sources, F2M~J0825 and F2M~J1532, have only narrow H$\beta$ lines visible in their optical spectrum, likely because the broad component has experienced significant extinction from dust. The H$\alpha$ line is shifted into a noisy part of the optical and near-infrared spectra precluding our ability to perform reliable Gaussian fitting. 
Both sources exhibit broad Pa$\beta$ emission in their near-infrared spectrum and their $M_{BH}$ values were computed in \citet{Kim15} along with 14 other red quasars using a single-epoch relation derived by \citet{Kim10} for Paschen lines,
\begin{equation}
\log \bigg(\frac{M_{\rm BH,vir}}{M_\odot} \bigg) = a + b \log \bigg(\frac{L_{{\rm Pa}\beta}}{10^{42} \rm erg/s} \bigg) + c \log \bigg(\frac{v_{\rm FWHM}}{1000 \rm km/s} \bigg),
\label{eqn:mbh_pa}
\end{equation}
where $a=7.04$, $b=0.48$, and $c=2$. 
\citet{Kim10} calibrated this relation using near-infrared spectra of unreddened quasars and found that they agree with the Balmer-line-based relations to within 0.18-0.24 dex.

In addition to line widths, the $M_{BH}$ relations require a luminosity at a particular wavelength to estimate the radial distance to the broad line region. 
The bolometric luminosity listed in Tables \ref{tab:previous_sources} and \ref{tab:sources} is determined by applying a bolometric correction of 7.6 to the 6$\mu$m luminosity based on the mean quasar spectral energy distribution (SED) in \citet{Richards06}. 
Because the luminosities in the $M_{BH}$ relations are at optical and UV wavelengths, which are affected by reddening in these quasars, we interpolate their Wide-field Infrared Survey Explorer \citep[{\em AllWISE};][]{Wright10,Mainzer11} mid-infrared fluxes 
to estimate their rest-frame 6$\mu$m luminosity, $L_{6\mu{\rm m}}$, and scale it to the optical flux using a ratio of the bolometric corrections for 6$\mu$m (7.6) and 5100\AA\ (10) for the \citet{Richards06} mean quasar spectral energy distribution (SED). 
To determine the uncertainty on $L_{6\mu{\rm m}}$, each SED is perturbed by its photometric errors, drawing from a Gaussian distribution, to generate 1000 SEDs which we interpolate to measure $L_{6\mu{\rm m}}$. The reported uncertainty is then the standard deviation of the $L_{6\mu{\rm m}}$ distribution.
$L_{6\mu{\rm m}}$ is also used to find the bolometric luminosity ($L_{\rm bol}$) of the quasars, applying a bolometric correction factor of 7.6 derived from the same SED. 
When combined with their bolometric luminosities, we are able to compute an Eddington ratio ($\lambda_{\rm Edd}$) for each source. 
Table \ref{tab:sources} lists the quasars, their positions, redshifts, and $E(B-V)$. The table also lists the source of the imaging, $M_{BH}$, $L_{\rm bol}$, and $\lambda_{\rm Edd}$.

\begin{table*}
\caption{Properties of Newly Added F2M Red Quasars}
\label{tab:sources}
\begin{tabular}{ccccccccccc}
\hline
{Name} & {R.A.} & {Decl.} & {Redshift} & {$E(B-V)$} & {$M_{BH}$} & {Line} & {$L/L_{\rm Edd}$} & {$\log{L_{\rm bol}}^\sharp$} & {Merger?} & {Image Ref} \\ 
{} & {(J2000)} & {(J2000)} & {} & {(mag)} & {($10^8 M_\odot$)} & {} & {} & {(erg s$^{-1}$)} & {} & {} \\
\hline
F2M~J0825{$^a$} & 08:25:02.00  &   +47:16:52.0  & 0.804 &  $0.68\pm0.01$  &  12.25$\pm$2.98  &  Pa$\beta$  & $0.67\pm0.18$ & $47.353\pm0.007$ & Y & U08 \\
F2M~J0834\phantom{a} & 08:34:07.00  &   +35:06:01.8  & 0.470 &  $0.91\pm0.02$  &  31$\pm$10{$^\dagger$} &  H$\alpha$  & $0.04\pm0.01$ &  $46.18\pm0.01$  & N & U08 \\
F2M~J1106\phantom{a} & 11:06:48.30  &   +48:07:12.3  & 0.435 & $0.443\pm0.002$ &  8.2$\pm$2.3           &  H$\alpha$  & $0.52\pm0.15$ & $46.734\pm0.005$ & ? & S23 \\
F2M~J1118\phantom{a} & 11:18:11.10  & $-$00:33:41.9  & 0.686 &  $0.61\pm0.01$  &  6.4$\pm$1.8.          &  H$\alpha$  &  $1.1\pm0.3$  & $46.943\pm0.008$ & Y & U08 \\
F2M~J1151\phantom{a} & 11:51:24.10  &   +53:59:57.4  & 0.780 &  $0.67\pm0.01$  &  5.86$\pm$0.04         &  H$\beta$   & $0.50\pm0.02$ &  $46.57\pm0.02$  & N & U08 \\
F2M~J1324\phantom{a} & 13:24:19.90  &   +05:37:05.0  & 0.205 & $0.326\pm0.003$ & 5.5$\pm$1.5{$^\ddagger$} & H$\alpha$  & $0.76\pm0.21$ & $46.718\pm0.005$ & Y & HST archive \\
F2M~J1507\phantom{a} & 15:07:18.10  &   +31:29:42.3  & 0.988 & $0.644\pm0.003$ &  4.6$\pm$1.3{$^\star$} &  H$\alpha$  & $1.48\pm0.42$ &  $46.93\pm0.01$  & Y & U08 \\
F2M~J1531\phantom{a} & 15:31:50.47  &   +24:23:17.6  & 2.287 & $0.311\pm0.004$ &  80$\pm$22.            &  H$\alpha$  & $1.61\pm0.46$ &  $48.21\pm0.02$  & Y & G15 \\
F2M~J1532{$^a$} & 15:32:33.19  &   +24:15:26.8  & 0.564 &  $0.68\pm0.03$  &  6.12$\pm$4.87   &  Pa$\beta$  & $0.29\pm0.27$ & $46.586\pm0.007$ & Y & U08 \\
F2M~J1715\phantom{a} & 17:15:59.80  &   +28:07:16.8  & 0.523 & $0.786\pm0.003$ &  8.7$\pm$2.4           &  H$\alpha$  & $0.33\pm0.09$ & $46.553\pm0.008$ & N & M03   \\
\hline
\multicolumn{11}{l}{$^\sharp$ Bolometric luminosities were determined by applying a bolometric correction of 7.6 \citep{Richards06} to the 6$\mu$m luminosity which was determined }\\
\multicolumn{11}{l}{\phantom{a} by interpolating the WISE mid-infrared luminosities in the rest-frame. }\\
\multicolumn{11}{l}{$^\dagger$ This object has a double-peaked emission line shape, resulting in a likely over-estimated $M_{BH}$ } \\ 
\multicolumn{11}{l}{$^\ddagger$ This source was fit with multiple Gaussian components to account for the presence of narrow lines.}\\
\multicolumn{11}{l}{$^\star$ This source has a blue-shifted broad line, which we do not use in our estimate of $M_{BH}$.(see \S \ref{appendix:f2m1507}).}\\
\multicolumn{11}{l}{$^a$ $M_{BH}$ and $L/L_{\rm Edd}$ were determined in \citet{Kim15}.}\\
\multicolumn{11}{l}{Comments -- M03 = \citet{Marble03}; U08 = \citet{Urrutia08}; S23 = \citet{Shen23}; U12 = \citet{Urrutia12};G15 = \citet{Glikman15}}\\
\end{tabular}
\end{table*}

\subsection{Chandra Observations} \label{sec:chandra}

We obtained {\em Chandra} observations in Cycle 21 of eight red quasars that obeyed our selection requirements outlined in Section \ref{sec:sel} that had no archival X-ray data (GO 21700216, PI: Glikman). 
We designed our observing strategy aiming for 70 counts in the {\em Chandra} energy range, which we estimated using the interpolated 6$\mu$m luminosity and the $L_{X} - L_{IR}$ relation, modified to reflect the trends seen for red quasars in \citetalias{LaMassa16b} and \citetalias{Glikman17a} (i.e., $\sim 1$ dex below the \citealt{Chen17} relation which accounts for any intrinsic $N_H$; see \S \ref{sec:lx_lir}), imposing a 5 ksec minimum on the brighter sources.  
Table \ref{tab:obs} lists the details of the {\em Chandra} observations for the eight sources as well as the two sources with archival observations. 
Given that the scatter in the $L_{X} - L_{IR}$ relation is on a logarithmic scale, while photon detection rates are linear, our total counts vary significantly from the expected 70. 

We processed the data with the CIAO v4.15, with CALDB v4.10.4 \citep{Fruscione06}, using the {\tt chandra\_repro} task to produce a filtered events file, removing periods of anomalously high background. 
For all but one of the observations, a spectrum was extracted using a 5\arcsec\ radius aperture around the object using the CIAO tool {\tt specextract}, with the background extracted from an annulus around the quasar with inner radius 10\arcsec\ and outer radius 35\arcsec. 
F2M~J1532 was present in two archival observations.
One of the archival observations had F2M~J1532 near the edge of the I2 chip on the ACIS-I detector where the PSF is significantly larger; we use a 35\arcsec\ radius aperture around the source and an offset circular aperture with a 120\arcsec\ radius far from any sources for the background. 
The total net counts detected are listed in Table \ref{tab:obs} as reported by the CIAO task {\tt dmlist}.

\begin{table*}
\caption{Summary of Chandra Observations}
\label{tab:obs}
\begin{tabular}{cclccc}
\hline
{Name} & {ObsID} & {Date} & {$N_{\rm H, Galactic}$} & {Net Exposure Time} & {Net Counts} \\ 
{} & {} & {} & {(10$^{20}$ cm$^{-2}$)} & {(ksec)} & {(0.5 - 7 keV cnts)} \\
\hline
F2M~J0825  & 22570 & 2019 December 20  & 4.15 & 9.94  & 218$\pm$15 \\
F2M~J0834  & 22571 & 2019 December 15  & 4.03 & 9.94  &  4$\pm$2  \\
F2M~J1106  & 22572 & 2019 October 24   & 1.38 & 5.99  & 2$\pm$2 \\
F2M~J1118  & 22573 & 2020 January 21  & 4.43 & 11.91 & 50$\pm$7 \\
F2M~J1151  & 22574 & 2020 August 7     & 1.33 & 17.83 & 22$\pm$5 \\
F2M~J1324  & 22575 & 2020 January 21   & 2.32 & 5.0   & 7$\pm$3 \\
F2M~J1507  & 22576 & 2019 November 9   & 1.66 & 19.80 & 70$\pm$9 \\
F2M~J1531  &  3336 & 2002 September 25 & 3.61 & 5.06  & 1$\pm$1 \\
F2M~J1532  &  ~3138{$^\dagger$} & 2001 April 30 & 4.14 & 47.13 & 441$\pm$24 \\
\ldots     &  3338 & 2002 July 2     & \ldots & 4.90  & 57$\pm$8 \\
F2M~J1715  & 22577 & 2019 October 5    & 3.79 & 8.95  & 255$\pm$16 \\
\hline
\multicolumn{6}{l}{$^\dagger$ Due to being far off-axis, this source was extracted from a 35\arcsec aperture and a nearby 120\arcsec-radius circular aperture for the background.}\\
\end{tabular}
\end{table*}

\section{X-ray fitting}

\subsection{Basic fits}\label{sec:xspec}

We perform spectral analysis only on sources with $>50$ counts.
Three sources are well detected with $>100$ counts which we grouped by a minimum of 5 counts per bin. Another two sources have between 50 and 100 counts, which we group by 2 counts per bin. 
We use the X-ray fitting software XSpec v12.13.0 \citep{Arnaud96} to model these sources. 
We use the Cash statistics \citep[C-stat;][]{Cash79} with direct background subtraction \citep{WLK79}. 

We began by fitting a simple power-law model, 
\begin{equation}
    {\tt phabs*zpowerlw}, \label{eqn:spl}
\end{equation}
allowing only absorption from gas in the Milky Way ({\tt phabs}).
Table \ref{tab:obs} lists the Galactic hydrogen column density, determined using the {\tt colden} CIAO task, which we freeze in all our fits. 
Given that red quasars experience absorption at optical wavelengths, we further fit an absorbed power-law model,
\begin{equation}
{\tt phabs*zphabs*zpowerlw}, \label{eqn:apl}
\end{equation}
with absorption occurring both at the source ({\tt zphabs}) and in the Milky Way to look for potential intrinsic obscuration in the source.
Finally, because the previous analyses of the X-ray spectra of red quasars revealed an excess of soft X-ray flux below 2 keV suggesting that there may be scattered or leaked light at lower energies in excess of the absorbed primary continuum \citepalias{LaMassa16b,Glikman17a}, we fit a double-absorbed power law with the same photon index for both components, 
\begin{equation}
{\tt phabs*(zpowerlw + zphabs*zpowerlw)}.    \label{eqn:dpl}
\end{equation}
We use an F-test to decide whether the additional components significantly improves the fit with a probability of $>95\%$. 
We report in Table \ref{tab:spec_fits} the fitted parameters for these sources, indicating in the second column the model equation used in the best fit. 

\subsection{Complex fits}
For one source, F2M~1507, there appears to be a reduction in flux around $4-5$ keV, which may be due to blue-shifted absorption from an outflow. We discuss the unusual spectral properties and perform more detailed fitting of this object to account for this absorption in Appendix \ref{appendix:f2m1507}.

In another source, F2M~1532, we noted the presence of an Fe K$\alpha$ line suggestive of reflection off a distant medium. 
While such emission is typically seen in Type 2 AGN, where the reflection occurs off of the obscuring torus, it has been seen in at least one red quasar \citep[F2M~0830;][]{Piconcelli10,LaMassa16b}, where the scattering may be due to clouds farther out from the nucleus. 
To address such scenarios, we turn to the MYTorus model of \citet{Murphy09} which solved the radiative transfer of X-rays from an AGN including scattering off of a torus, line-of-sight absorption, as well as leakage or scattered light. 
In XSpec, the model is defined similar to Eqn \ref{eqn:dpl}:

\begin{multline}
C \times {\tt phabs} \times [{\tt zpowerlw} \times {\tt MYTorusZ(N_{H,Z},\theta_{\rm obs}, E)} \\
+ A_S \times {\tt MYTorusS(} \Gamma \tt{, N_{H,S},\theta_{\rm obs},E)} \\
+ A_L \times {\tt MYTorusL(} \Gamma \tt{, N_{H,S},\theta_{\rm obs},E)} \\
+ f_{\rm scatt} \times {\tt zpowerlw}].\\
\label{eqn:mytorus}
\end{multline}

where {\tt E} is the observed energy and {\tt MYTorusZ}, {\tt MYTorusS}, and {\tt MYTorusL} are tables that contain pre-calculated parameters derived via Monte Carlo calculations that take into account the reprocessing of the intrinsic AGN continuum in a toroidal structure for a range of column densities. 
{\tt MYTorusZ} is the so-called `zeroth-order spectrum', and represents the intrinsic continuum that makes it through any absorbing or scattering medium along the line-of-sight ({\em mytorus\_Ezero\_v00.fits}).  
{\tt MYTorusS} tabulates Compton-scattered emission that is added to the zeroth-order spectrum ({\em mytorus\_scatteredH500\_v00.fits}). 
{\tt MYTorusL} provides fluorescent line emission that is also added to the zeroth-order spectrum ({\em mytl\_V000010nEp000H500\_v00.fits}, where {\em H200} refers to the termination energy of the model of 200 keV).  
This model set up is the same as previously used for the analysis of F2M red quasars \citepalias{LaMassa16b,Glikman17a} as well as 3C~223, whose complex X-ray spectrum has characteristics similar to F2M~J1532 \citep{LaMassa23}.
All three MYTorus components are needed in order to preserve the self-consistency of the model.  
We discuss the detailed fitting of F2M~J1532 in Appendix \ref{appendix:f2m1532}.

We report the results of these complex fits in \ref{appendix:f2m1507} and \ref{appendix:f2m1532}. However, since the essential parameters ($\Gamma$, $N_H$, $f_{\rm scatt}$) used in the subsequent analysis are similar to the phenomenological results, we do not use the complex fitted parameters in the subsequent analysis. 

\begin{table*}
\caption{Best fit parameters for high count sources}
\label{tab:spec_fits}
\begin{tabular}{ccccccc}
 \hline
{Name} & {Model} & {$\Gamma$} & {$\log{N_{H}}$} & {$f_{\rm scatt}$} & {C-stat} & {HR} \\ 
{} & {Eqn.} & {} & {(cm$^{-2}$)} & {(\%)} & {(DOF)} & {} \\
\hline
F2M~J0825 & 2 & $2.38^{+0.57}_{-0.54}$ & $22.78^{+0.17}_{-0.24}$ & \ldots    & 44.49 (36) &    0.19  \\
F2M~J1118 & 2 & $2.18^{+1.18}_{-0.99}$ & $22.45^{+0.36}_{-1.85}$ & \ldots    & 10.96 (22) &  $-0.06$ \\
F2M~J1507 & 3 & $1.8${$^\sharp$}        & $23.5\pm0.4$           & 16        & 28.29 (27)  &    0.31 \\
F2M~J1532{$^\dagger$} & 3 & $1.3\pm0.5$ & $22.90^{+0.19}_{-0.33}$ & 11 & 113.48 (120) & 0.44,0.38{$^\ddagger$} \\
F2M~J1715 & 1 & $1.57\pm0.21$          & \ldots              & \ldots    & 57.72 (42)  &  $-0.08$  \\
\hline
\multicolumn{7}{l}{$^\sharp$ The photon index for this fit was fixed due to the small number of bins. (See \S \ref{appendix:f2m1507} for a more complex modeling approach). }\\
\multicolumn{7}{l}{$^\dagger$This source was fit jointly with both observations listed in Table \ref{tab:obs}.}\\
\multicolumn{7}{l}{$^\ddagger$ The HRs were computed separately for each of the observations listed in Table \ref{tab:obs}, with the longer observation (ObsID 3138) listed first.}\\
\end{tabular}
\end{table*}

\begin{figure*}
\centering
\includegraphics[scale=0.31]{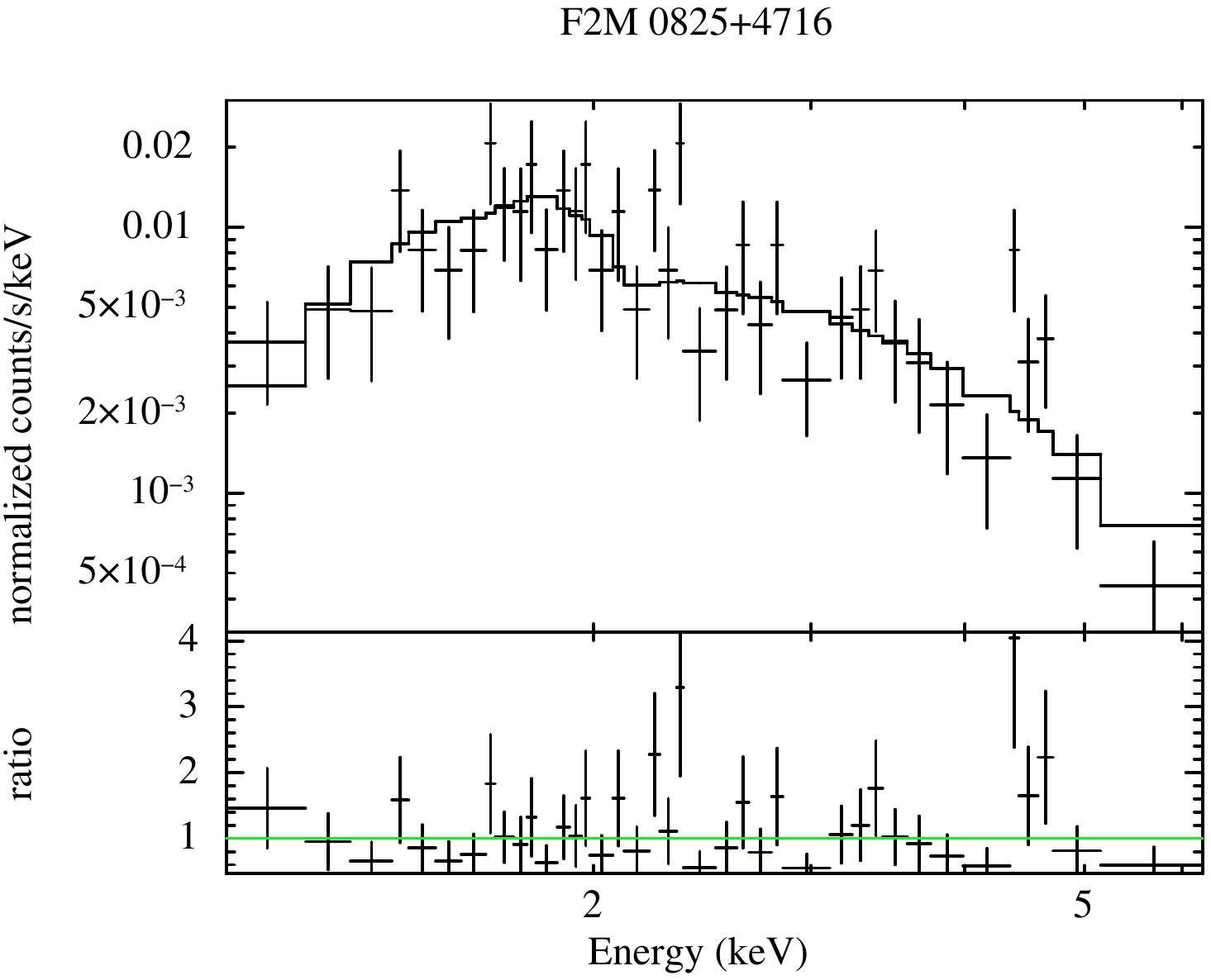}
\includegraphics[scale=0.31]{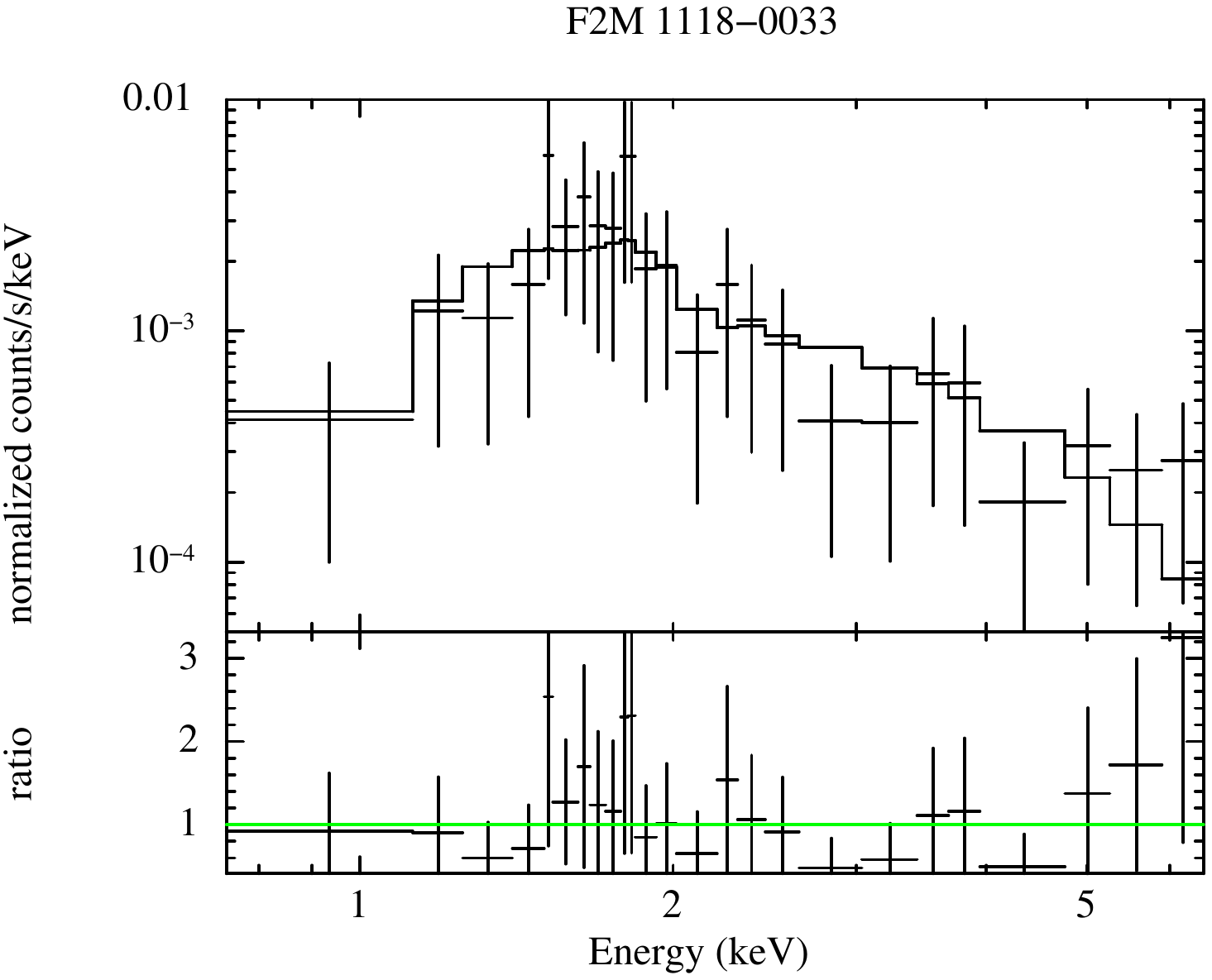}
\includegraphics[scale=0.31]{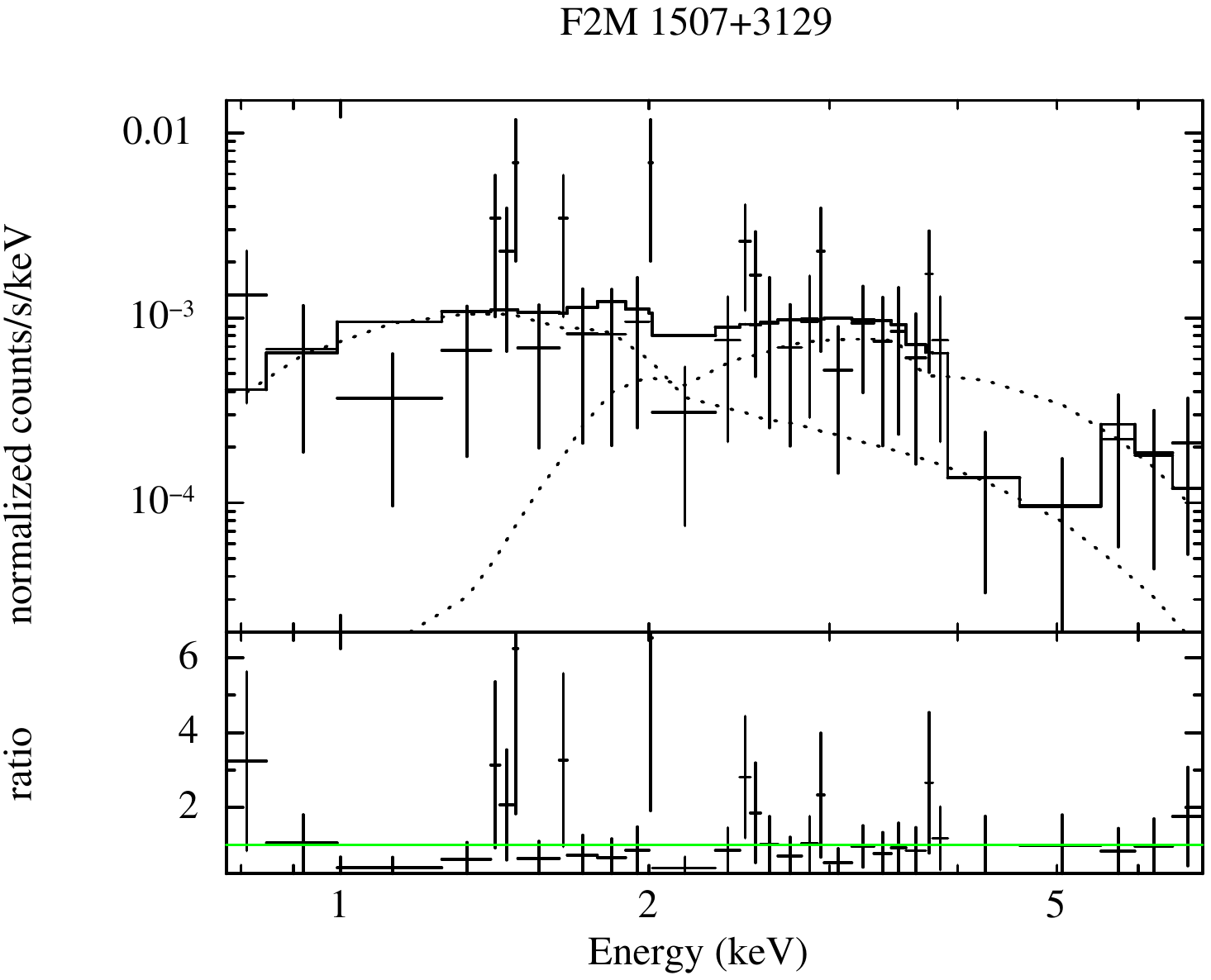}
\includegraphics[scale=0.31]{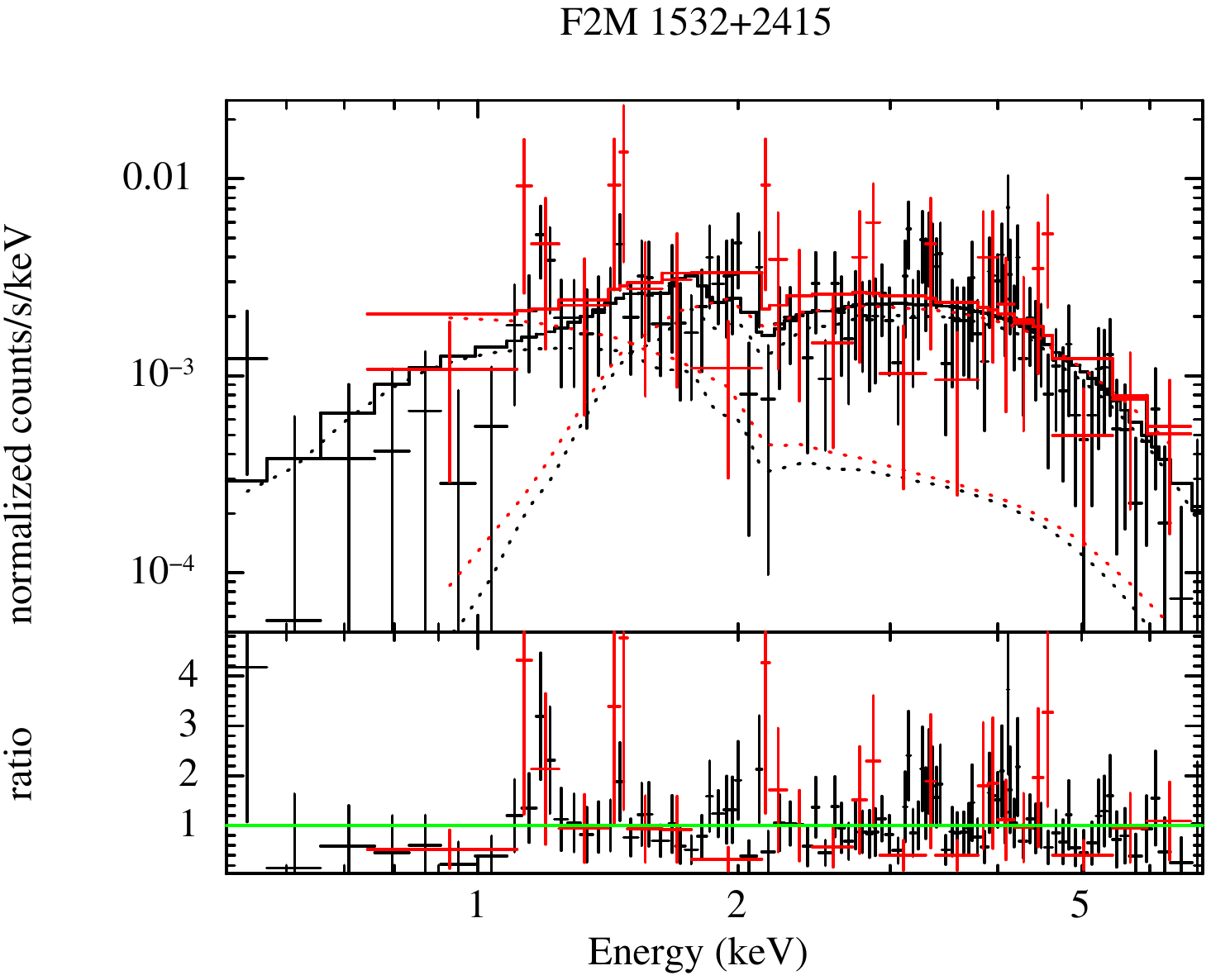}
\includegraphics[scale=0.31]{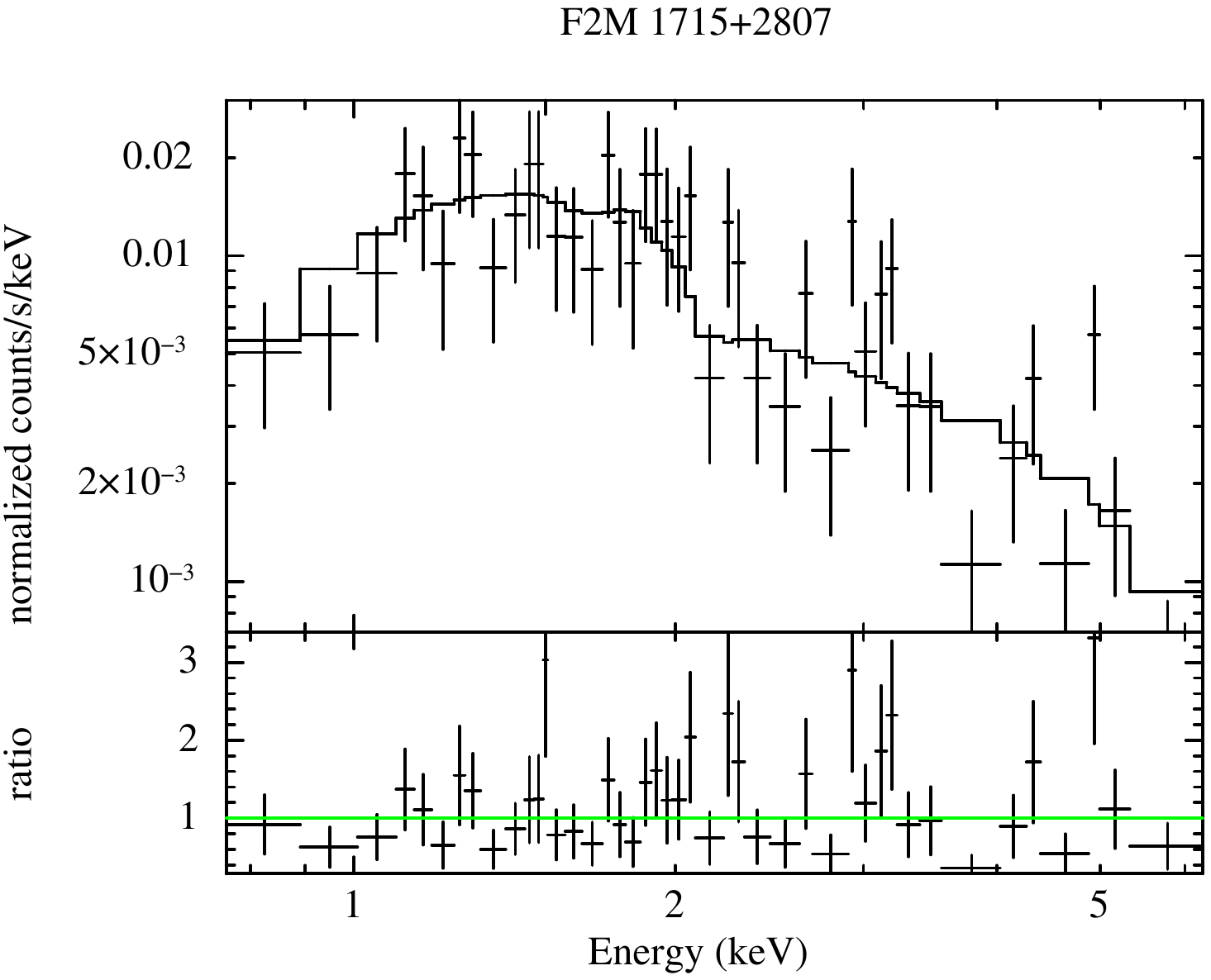}
\caption{Best model fits to the X-ray spectra for counts in the energy range 0.5 -- 7 keV, as described in Table \ref{tab:spec_fits}. Data are shown as points with error bars. The solid lines represent the best-fit model with dotted lines representing the individual components of a partially-covered model (Eqn. \ref{eqn:dpl}), when applicable. The bottom panels of each figure show the counts-to-model ratios. For F2M~J1532, the black and red points represent the two archival data sets used for the fitting (ObsID 3138 and ObsID 3338, respectively). The best fit models are coloured correspondingly. }
\label{fig:xray_spectra}
\end{figure*}

\subsection{Hardness ratios in the low count regime}

For the three sources with $\gtrsim 5$ and $\lesssim 50$ counts -- an insufficient amount for spectral modeling -- we instead report hardness ratios (HRs), which are a meaningful proxy for X-ray absorption. 
The HR is defined by comparing the net counts in the hard ($H$) and soft ($S$) bands, defined as $0.5 - 2$~keV and $2 - 7$~keV in the observed frame, respectively, as appropriate for {\em Chandra}'s energy response, by the expression $(H - S)/(H + S)$.  
We determine these counts via the CIAO command {\tt dmcopy} which filters the events file to create an image with just the photons in each energy band. We then apply the same source and background regions to measure the source counts using the CIAO tool {\tt dmextract}. 

Given that we are in this low-count regime, we employ the Bayesian Estimation of Hardness Ratios \citep[BEHR;][]{Park06} code which determines HRs, properly handling uncertainties in the Poisson limit, including non-detections. 
We report in Table \ref{tab:hr} the hard and soft counts as well as the mode of the HR determined by BEHR. The stated uncertainties represent the lower and upper bounds reported by BEHR. 

Assuming an absorbed power-law model for these red quasars (Eqn. \ref{eqn:apl}) and fixing the power-law index to $\Gamma = 1.8$, we can crudely approximate the column density responsible for the measured HR. 
Following a similar approach described in \citet{Martocchia17}, we simulate such a spectrum with the WebPIMMS interface\footnote{\tt https://cxc.harvard.edu/toolkit/pimms.jsp} setting the appropriate Cycle of the observations, providing the soft count rate for each quasar, varying the intrinsic $N_H$, and computing the HR from the predicted hard count rate until the lower bound reported by BEHR is reached. 
We then regard the $N_H$ value from the simulated spectrum as representing a lower-limit for the absorption. 
We report these values in Table \ref{tab:hr} as well. 
We note that the $N_H$ values derived from HRs are highly simplified, as they neglect scattering or leakage of X-ray photons (e.g., Eqn. \ref{eqn:dpl}) and assume a fixed power-law continuum, $\Gamma$. 

Because they are computed in the observed band, HRs depend on redshift, with the strongest dependence occurring for moderately absorbed sources ($10^{22} < N_H < 10^{23}$ cm$^{-2}$) at $z<1$ \citep{LaMassa16a} which is where all the quasars with computed HRs in this paper lie. 
None the less, in all cases, HRs $\gtrsim 0$ imply $N_H \gtrsim 10^{22}$ cm$^{-2}$. Mindful of these considerations, we compare these values to HRs measured in other red quasar samples.

\citet{Glikman18} surveyed the $270$ deg$^2$ equatorial region known as SDSS Stripe 82 \citep{Frieman08} which contains a wealth of multi-wavelength ancillary data.
Using near-to-mid infrared selection to a relatively shallow flux limit of 20 mJy at 22$~\mu$m, they identified 21 red QSOs, most lacking a radio detection in FIRST but with otherwise similar characteristics as the F2M red quasars. 
Four red QSOs in that study had X-ray detections that allowed for HR measurements. Their redshifts span $z = 0.2$ to $z = 0.83$ and HR $= -0.085$ to HR $= 0.863$, with the $z=0.200$ object having HR = 0.792, thus placing them all in the moderately absorbed ($N_H > 10^{22}$ cm$^{-2}$) regime. 

In an X-ray selected red QSO sample over Stripe 82, reaching significantly fainter sources (including SDSS drop-outs) and thus higher redshifts up to $z=2.5$, \citet{LaMassa17} find 12 sources displaying features consistent with the evolutionary paradigm proposed for the F2M red quasars. 
\citet{LaMassa17} measure a range of HRs and, applying a similar translation between HR and $N_H$, find that half have $N_H > 10^{22}$ cm$^{-2}$ with three sources consistent with no absorption. 
The same caveats about soft excess due to scattering and leakage apply here as well such that higher-count X-ray spectra may reveal more complex physics than a simple absorbed power law. 

\begin{table*}
\caption{Hardness ratios and absorption in low-count sources}
\label{tab:hr}

\begin{tabular}{ccccc}
\hline
{Name} & {Net Soft } & {Net Hard} & {HR} & {$\log(N_H)$} \\ 
{} & {(counts)} & {(counts)} & {} & {(cm$^{-2}$)} \\
\hline
F2M~J1106 & $<0.07$ &  $2.33\pm1.74$ & $0.99_{-0.39}^{+0.01}$ & $>22.9$ \\
F2M~J1151 &  $5.91\pm2.65$ & $16.31\pm4.25$ & $0.49_{-0.21}^{+0.19}$ & $>22.8$ \\
F2M~J1324 &  $2.69\pm1.73$ &  $4.44\pm2.24$ & $0.29\pm0.38$ & 21.7 (22.4){$^\dagger$} \\
\hline
\multicolumn{5}{l}{$^\dagger$ This source's lower and upper bound spanned a very broad range; we provide in parentheses the $N_H$ corresponding to the mode value.}\\
\end{tabular}
\end{table*}

\subsection{Upper limits for undetected sources}

Two sources, F2M~J0834 and F2M~J1531, have counts consistent with non-detections. 
For these sources, we follow the CIAO thread for calculating source count rates and model-independent fluxes. We compute the flux over the full energy range using the task {\tt srcflux} modeling the flux as being absorbed by Milky Way gas ({\tt phabs}). 
We note that F2M~J1531 is the only high redshift source in this sample, having $z=2.287$ while the rest are all at $z<1$, and is thus the only one with imaging from \citet{Glikman15}.
Therefore, although its morphology shows evidence of a merger, its heterogeneous imaging aspects do not impact the results presented in Section \ref{sec:results}.
 \\

Having extracted flux information from all ten sources, we present the soft (0.5–2 keV), hard (2–10 keV), and full (0.5–10 keV) X-ray fluxes in Table \ref{tab:fluxes}\footnote{Although the hard band was defined as $2-7$ keV when computing HRs, we define the hard band as $2-10$ keV when reporting fluxes and luminosities so we can compare them  with established X-ray relations in the literature that use that band definition.}. 
We also compute the X-ray luminosities in the 2-10 keV band, which are used to compare with other emission diagnostics in Section \ref{sec:results}. 
For objects with sufficient counts to enable spectral fitting (i.e., those listed in Table \ref{tab:spec_fits}) we measure and report the observed luminosity using the best-fit model; we omit Milky Way absorption in this calculation. 
We then also report an absorption-corrected luminosity by defining a simple {\tt zpow} model with the best-fit power-law index ($\Gamma$) and its normalization. The uncertainties on the luminosity are derived from the uncertainty on the power-law normalization. 
For the low count objects (i.e., those listed in Table \ref{tab:hr}), we determine their luminosities assuming a power-law spectrum ({\tt zpow}) with an index of $\Gamma = 1.8$. 
We normalize this model based on a fit to the low count data using the model in Eqn \ref{eqn:apl}, and derive the uncertainties on the luminosity from the uncertainties on the power-law normalization in this model. 
Given that $N_H$ for these sources was estimated from the HRs, which are already uncertain, we do not compute a luminosity from the observed data.
We do not compute a luminosity for the two sources that we deem to be undetected and for which we report upper limits to their fluxes.  
Table \ref{tab:fluxes} also reports the rest-frame absorption-corrected 2-10 keV luminosities as well as their rest-frame $6\mu$m luminosities, which are determined by interpolating between the {\em WISE} photometric bands, as described in Section \ref{sec:sel}. 

\begin{table*}
\caption{Observed X-ray Fluxes}
\label{tab:fluxes}
\begin{tabular}{cccccc}
\hline
{Name} & {$F_{0.5-2~{\rm keV}}$} & {$F_{2-10~{\rm keV}}$} & {$F_{0.5-10~{\rm keV}}$} & {$\log L_{2-10~{\rm keV, int}}$} & {$\log L_{6~\mu{\rm m}}$} \\ 
{} & {($10^{-14}$ erg cm$^{-2}$ s$^{-1}$)} & {($10^{-14}$ erg cm$^{-2}$ s$^{-1}$)} & {($10^{-14}$ erg cm$^{-2}$ s$^{-1}$)} & {(erg s$^{-1}$)} & {(erg s$^{-1}$)} \\
\hline
F2M~J0825 &  $5.4_{-4.2}^{0.5}$      &  $27.9_{-18.3}^{+0.7}$   &  $33.3_{-32.8}^{+1.4}$ & $45.10^{+0.46}_{-0.44}$            & 46.472$\pm$0.007 \\ 
F2M~J0834 &  \ldots                  &  \ldots                  &  $<1.1${$^\dagger$}    & \ldots                             &  45.30$\pm$0.01 \\ 
F2M~J1106 &  $<0.0003$               &  $2.4_{-1.9}^{+1.5}$     &  $2.4_{-2.2}^{+1.3}$   & $43.58^{+0.45}_{-0.80}$$^\ddagger$ & 45.854$\pm$0.005 \\ 
F2M~J1118 &  $<1.4$                  &  $4.9_{-4.5}^{+0.1}$     &  $<6.2$                & $44.08^{+0.81}$                    &  46.062$\pm$0.008 \\ 
F2M~J1151 &  $0.11_{-0.03}^{+0.02}$  &  $2.8_{-0.7}^{+0.8}$     &  $2.9_{-0.7}^{+0.6}$   & $43.96^{+0.15}_{-0.18}$$^\ddagger$ &  45.69$\pm$0.02 \\ 
F2M~J1324 &  $0.22_{0.09}^{+0.1}$    &  $2.5_{-1.0}^{+1.3}$     &  $2.7_{-1.4}^{+1.2}$   & $42.53^{+0.27}_{-0.37}$$^\ddagger$ & 45.837$\pm$0.005 \\ 
F2M~J1507 &  $1.1_{-0.6}^{+0.2}$     &  $7.3_{-1.3}^{+2.6}$     &  $8.4_{-2.9}^{+1.2}$   & $44.46^{+0.21}_{-0.29}$            &  46.05$\pm$0.01 \\ 
F2M~J1531 &  \ldots                  &  \ldots                  &  $<0.6${$^\dagger$}    & \ldots                             &  47.33$\pm$0.02 \\ 
F2M~J1532 &  $2.0_{-0.4}^{+0.2}$     &  $36_{-17}^{+1}$         &  $38.0_{-13}^{+1}$     & $44.56^{+0.46}_{-0.49}$            & 45.706$\pm$0.007 \\ 
F2M~J1715 &  $13.9_{-1.5}^{+1.6}$    &  $35_{-6}^{+4}$          &  $48.0_{-3.7}^{+3.3}$  & $44.48\pm0.11$                     & 45.673$\pm$0.008 \\ 
\hline
\multicolumn{6}{l}{$^\dagger$ These fluxes are reported over the 0.5-7 keV range as they are derived directly from the data with the {\tt srcflux} task on undetected sources.}\\
\multicolumn{6}{l}{Note -- Upper limits are quoted when XSpec returns a 1-$\sigma$ lower limit of 0.}\\
\multicolumn{6}{l}{$^\ddagger$ These sources had too few counts for spectral modeling. Their intrinsic luminosities are modeled from a fixed $\Gamma = 1.8$}\\
\end{tabular}
\end{table*}

\section{Results and Discussion} \label{sec:results}

Following the definition and identification of F2M red quasars as a population in \citet{Glikman04}, several other reddened and obscured quasar samples have been constructed using various definitions that exhibit similar characteristics of being in a transitional phase of quasar evolution. 
Many of these samples' selection criteria overlap the F2M selection, but extend along other parametric axes. 
We summarize here the various reddened AGN populations and compare their X-ray-derived properties to the F2M sample in this work. 

F2M red quasars were selected by applying the optical to near-infrared colour cuts of $(R - K)_{\rm Vega}> 4$ mag and $(J - K)_{\rm Vega} > 1.7$ mag to sources with matches in FIRST and 2MASS. 
An essential selection criterion for F2M red quasars is that they exhibit at least one broad ($v_{\rm FWHM} > 1000$ km s$^{-1}$) emission line in their spectrum; therefore F2M red quasars are by definition Type 1 sources. 
Although the $J-K$ colour cut avoids most low mass (M class) stars, they remain a strong contaminant since they are abundant in the Galaxy and have colours that resemble reddened quasars \citep[c.f.,][]{Warren00}. 
Radio selection was invoked to more thoroughly avoid them, but as a result the F2M survey misses large numbers of radio-faint red quasars. 

\citet{Banerji12} and \citet{Temple19} invoked a more stringent $(J - K)_{\rm Vega} > 2.5$ mag colour cut which naturally identifies more heavily reddened systems at higher redshifts ($z \gtrsim 1.5$). 
The sample is restricted to broad line (Type 1) sources and consists of $\sim 50$ objects. 
These heavily reddened quasars (HRQs) are also intrinsically more luminous and show outflows in [\ion{O}{iii}]. 
Although no rest-frame high-resolution optical imaging exists to identify whether the HRQs reside in merging hosts, an ALMA observation of one HRQ, J2315, does show merging evidence \citep{Banerji21}.

Aiming to exploit the mid-infrared photometry from {\em WISE}, which is less sensitive to dust extinction, a population of hyperluminous, hot dust obscured galaxies \citep[Hot DOGs;][]{Wu12,Tsai15} was identified by the ``W1W2 dropout'' method such that they are weak or undetected at 3.4 $\mu$m and 4.6 $\mu$m but bright at 12 $\mu$m and 22 $\mu$m. 
There are only $\sim 1000$ such sources across the entire extragalactic sky and their redshifts are in the cosmic noon era ($z\simeq 2-4$).
These objects likely contain buried AGN whose presence is implied by hot dust temperatures $\sim 60 - 120$ K and optical spectroscopic diagnostic features, though broad lines are often not seen.  
\citet{Fan16} investigated the morphologies of 18 Hot DOGs with HST imaging and concluded a merger fraction of $62\pm14$\% which is lower than for the F2M red quasars ($>80$\%) but higher than unobscured AGN hosts \citep[$\sim30$\%;][]{Villforth23}.
\citet{Farrah17} find a similar merger fraction ($\sim 75\%$) in their HST study of Hot DOGs but conclude that this high fraction is reflective of the massive galaxy population at $z\sim 2$.
Hot DOGs have been rarely detected in X-rays, likely due to the common presence of heavy ($N_H>10^{24}$ cm$^{-2}$) absorption \citep[e.g., ][]{Piconcelli15,Vito18}.
Hot DOGs are also interpreted as representing an evolutionary phase. 

`Extremely red quasars' (ERQs) were selected by the optical-to-mid-infrared colour $r_{\rm AB} - W4_{\rm Vega} > 14$ mag in \citet{Ross15} and $i_{\rm AB} - W3_{\rm Vega} > 9.8$ mag plus \ion{C}{iv} line properties indicative of outflows in \citet{Hamann17} resulting in $\gtrsim 300$ ERQs.
These criteria pick out objects that are more heavily reddened than the F2M red quasars at redshifts similar to the HRQs ($2 < z < 4$). ERQs contain Type 1 and Type 2 sources, the latter exhibiting significant amounts of polarization \citep{Alexandroff18,Zakamska23}. 
Their hosts are largely not in mergers \citep[only 2/10 sources studied show merger activity;][]{Zakamska19} but exhibit powerful winds seen in broad [\ion{O}{iii}] emission lines in excess of 1000 km s$^{-1}$ and with sufficient energy to impact their hosts \citep{Vayner21,Lau22}. 

Aiming to overcome the radio-selection of the F2M survey, and to exploit the wealth of multi-wavelength data in the SDSS Stripe 82 region, \citet{LaMassa16a} and \citet{Glikman18} used X-ray selection and {\em WISE} colours, respectively, to identify additional samples of red quasars. 

In addition, \citet{Jun21} performed a meta analysis of the aforementioned obscured AGN populations to relate their X-ray absorption and dust extinction properties. 
In the following sections, we compare the F2M quasars in this work to those compiled results and place F2M red quasars in the broader context of luminous obscured quasars. 

\subsection{Dust-to-gas ratios}\label{sec:d2g}

The X-ray data for the F2M red quasars presented here provide a measure of the column density, $N_H$, which parametrizes the absorption due to atomic gas along the line of sight. 
This value is determined via spectral fitting for the five sources with $\gtrsim 50$ counts and is considered more reliable than the $N_H$ estimated from the HR measured for sources fewer counts. 
The dust extinction, parametrized by $E(B-V)$, is reported in Tables \ref{tab:previous_sources} and \ref{tab:sources}. 
Together, $E(B-V)$ and $N_H$ provide constraints on the nature of the absorber, namely its dust-to-gas ratio. 

\begin{figure}
\includegraphics[scale=0.6]{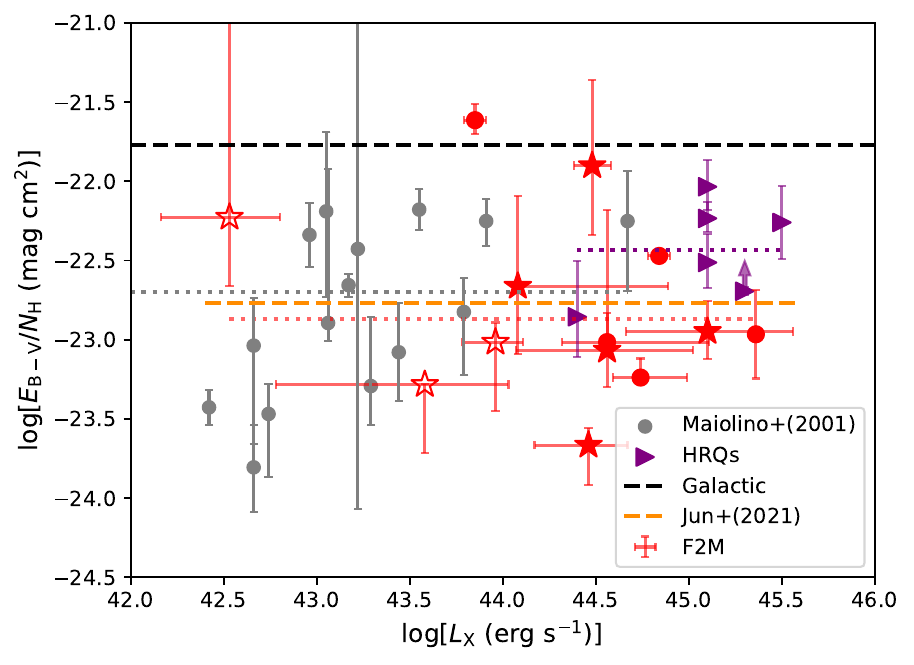}
\caption{Dust-to-gas ratios versus 2-10 keV X-ray luminosity ($L_X$). 
The Milky Way dust-to-gas value is depicted with a black dashed line.
The five previously studied quasars from \citetalias{LaMassa16b} and \citetalias{Glikman17a} and F2M~J0915 are shown with filled red circles. 
Filled red stars are the five sources in this work that had sufficient counts for spectral modeling. 
Open stars are the three low-count sources whose absorption was estimated from HRs. 
The horizontal red dotted line marks the mean value. 
For comparison, we plot the sample studied by \citet{Maiolino01} of low-luminosity AGN with gray circles and the mean value is depicted by a dotted gray line. 
We also show the HRQs from \citet{Lansbury20} with purple triangles and their mean value with a dotted purple line. 
The orange dashed line shows the mean dust-to-gas ratio of a compilation of red and obscured quasars in the literature \citep{Jun21}.
}
\label{fig:d2g}
\end{figure}

In Figure \ref{fig:d2g} we plot the dust-to-gas ratio for the F2M red quasars as a function of their 2-10 keV X-ray luminosity with red symbols, where filled circles are the four previously-studied quasars from \citetalias{LaMassa16b} and \citetalias{Glikman17a} as well as F2M~J0915 from \citet{Urrutia05}. 
Filled stars represent the five quasars that had $N_H$ determined via spectral fitting and open star symbols are the three sources whose $N_H$ values were estimated from their HRs and are therefore least precise. 

It has already been demonstrated in \citet{Maiolino01} that low-luminosity AGN (i.e., Seyfert galaxies) have dust-to-gas ratios that are significantly lower than the interstellar medium value determined for the Milky Way \citep[$1.7\times10^{-22}$ mag cm$^2$ shown with a horizontal black dashed line;][]{Bohlin78}. 
These Seyfert AGN were selected to have simple X-ray absorption spectra, avoiding sources with warm absorbers or cold absorbers with partial-covering (Eqn. \ref{eqn:dpl}). 
The dust-to-gas ratios found for these AGN, plotted with gray circles \ref{fig:d2g}, is therefore descriptive of circumnuclear material. The mean value for this sample is $\log{E(B-V)/N_H} = -22.8$, shown with a horizontal dotted gray line.

Given that most of the F2M red quasars are found in mergers, are fit by a variety of absorption models including partial covering, and are more luminous by $\sim 1-2$ orders of magnitude, they may not have the same source of reddening as the Seyferts in \citet{Maiolino01}. 
Yet, the dust-to-gas ratio distribution of F2M red quasars overlaps the \citet{Maiolino01} sample. 
Since the lower dust-to-gas ratio is suggestive of larger dust grains \citep{Maiolino01b}, but is also consistent with dust grains being sublimated close to the central engine, it is possible that different mechanisms produce similar results. 
The two sources whose dust-to-gas ratios are consistent with the Milky Way value are F2M~J1715, which had the lowest measured $N_H$ value such that its best-fit spectrum was a single power law with no absorption (Eqn. \ref{eqn:spl})\footnote{To compute a dust-to-gas ratio, we use the $N_H$ from an absorbed power-law fit (Eqn. \ref{eqn:apl}; $N_H = 6\times10^{21}$ cm$^{-2}$), which was a poorer fit than a simple unabsorbed power law according to an F-test, but allows for an estimate of the dust-to-gas ratio.}, and F2M~J1227 from \citetalias{LaMassa16b}, which lacks imaging to determine whether it is hosted by a merger and has minimal line-of-sight absorption ($N_H = 0.34\times10^{22}$ cm$^{-2}$).
The rest of the sources have lower dust-to-gas ratios than the Milky Way value by factors similar to the \citet{Maiolino01} sample. In addition, the mean dust-to-gas ratio is $\log{E(B-V)/N_H} = -22.9$ (dotted red line), which consistent with the previous studies in \citetalias{LaMassa16b} and \citetalias{Glikman17a}.

For comparison, we plot the average dust-to-gas ratio for six HRQs studied in \citet{Lansbury20} with purple triangles. Their mean value, $\log{E(B-V)/N_H} = -22.3$ shown with a dotted purple line, is higher than the F2M red quasars and the \citet{Maiolino01} sample. 
This may be explained by the more stringent colour selection of the HRQ sample which finds preferentially more reddened sources with higher $E(B-V)$ values. 
The meta analysis of \citet{Jun21}, which includes the previously published F2M red quasars, Type 2 AGN, ERQs, and Hot DOGs, finds a value consistent with the \citet{Maiolino01} average ($\log{E(B-V)/N_H} = -22.77\pm0.41$; dashed orange line). 

\subsection{X-ray versus infrared luminosity}\label{sec:lx_lir}

There exist various tracers of intrinsic AGN luminosity and, by extension, SMBH accretion that must be reconciled in order to adopt a consistent physical model for AGN. X-rays trace the innermost emission from the accretion disk, while mid-infrared (e.g., $6~\mu$m) arises from reprocessed UV photons from the accretion disk that are absorbed by nuclear dust (i.e., the `torus') and thermally re-emitted. At low luminosities, X-ray and mid-IR emission follow a linear relation \citep[in log-log space;][]{Lutz04}. However, observations of higher-luminosity quasars show a departure from this relation around $L_{\rm bol} \sim 10^{44}$ erg s$^{-1}$ \citep{Stern15,Chen17} and are are interpreted as being under-luminous in X-rays \citep{Ricci17b}. 
The decreasing $L_X/L_{\rm IR}$ ratio with increasing luminosities could also be interpreted as due to the increasing bolometric correction in X-rays at high $L_{\rm bol}$ \citep{Martocchia17}.

\begin{figure}
\includegraphics[scale=0.58]{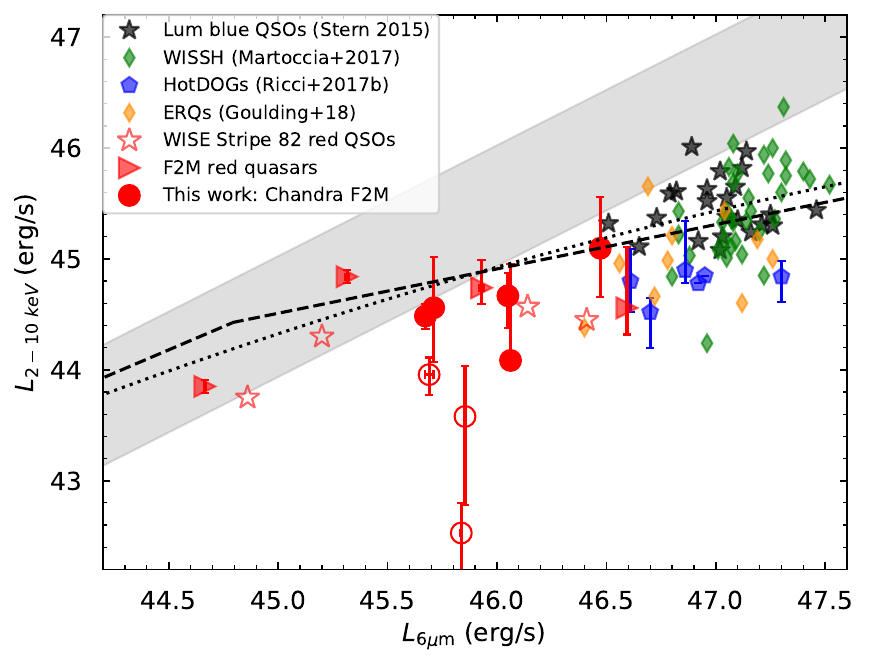}
\caption{Rest frame 6 $\mu$m luminosity vs. rest-frame absorption-corrected 2-10 keV X-ray luminosity for different quasar samples. 
Results from this work are shown with red circles. The six high-count sources that had successful spectral modeling are shown with filled circles, while the open circles are the three low-count sources that were modeled by a fixed $\Gamma=1.8$ power law. 
Other red quasars from F2M \citepalias{LaMassa16b,Glikman17a}  are shown with red triangles. 
Red stars are {\em WISE}-selected red quasars in Stripe 82 \citep{Glikman18}, which have not been corrected for absorption.
Apart from the two lowest flux sources whose luminosities may have been significantly underestimated due to insufficient absorption correction, the newly added F2M red quasars populate a similar part of this space as the previously studied red quasars. 
More luminous quasar samples are shown for comparison.
Black stars are unobscured, Type 1 quasars from \citet{Stern15}. The relation that was derived from those data is shown with a dotted line. 
Hyperluminous Type 1 quasars from the WISSH sample \citep{Martocchia17} are shown with green diamonds.
Asterisks show Hot DOGs \citep{Ricci17b}, which are infrared-hyperluminous, heavily obscured quasars. 
ERQs are shown with orange diamonds \citep{Goulding18}.
The shaded region shows the \citet{Lutz04} relation derived from local Seyfert galaxies which breaks down at high luminosities. The dashed line represents the relation from \citet{Chen17} derived from luminous AGN in deep fields.}
\label{fig:lx_lir}
\end{figure}

In Figure \ref{fig:lx_lir} we plot the eight F2M red quasars with X-ray luminosities (Table \ref{tab:fluxes}) along with other samples from the literature. 
The black stars and green asterisks are luminous, unobscured, Type 1 quasars from \citet{Stern15} and \citet{Martocchia17}.
The low luminosity relation from \citet{Lutz04}, which is calibrated at luminosities too low to appear on this plot, is shown by the shaded region. 
The relations from \citet{Stern15} and \citet{Chen17}, defined based on unobscured Type 1 samples, are shown with dotted and dashed lines, respectively, and depart from the shaded region. 
ERQs are shown with orange diamonds \citep{Goulding18} and exist in the same part of the $L_X - L_{\rm IR}$ relation as the unobscured objects. 
Hot DOGs, shown with blue pentagons, fall systematically below the extended relations \citep{Ricci17b}. 

Models of radiation-driven feedback postulate that X-rays need to be suppressed in order to enable line-driven winds on small scales \citep{Proga00}. 
And it has been shown that quasars with strong outflows are X-ray weak \citep{Luo13,Zappacosta20}. 

While the F2M red quasars (red symbols, both from this and previous works) are not as luminous in the infrared as the Hot DOGs or ERQs, they similarly lie below the $L_X - L_{\rm IR}$ relation established at low luminosities, even when corrected for absorption. 
F2M red quasars have an anomalously high fraction of BAL systems indicative of line-driven outflows. 
In addition, F2M~J1507 shows evidence for ultra-fast out-flowing material in its X-ray spectrum (\S \ref{appendix:f2m1507}).
In this space, Hot DOGs appear to be an extension of F2M red quasars toward more luminous IR sources whose X-rays are suppressed compared to their IR luminosity \citep[for a discussion on the X-raw weakness of Hot DOG, see][]{Ricci17b}. 

We note that the open F2M symbols were in the low-count regime and had their column density estimated from their HRs which was used to correct the X-ray luminosity. These luminosities are therefore highly uncertain and may be underestimated. However, given that their exposure times are similar to the rest of the sample, while their net counts are lower by more than an order of magnitude, they are likely intrinsically less luminous.

\subsection{Radiative feedback}

\begin{figure*}
\includegraphics[scale=0.58]{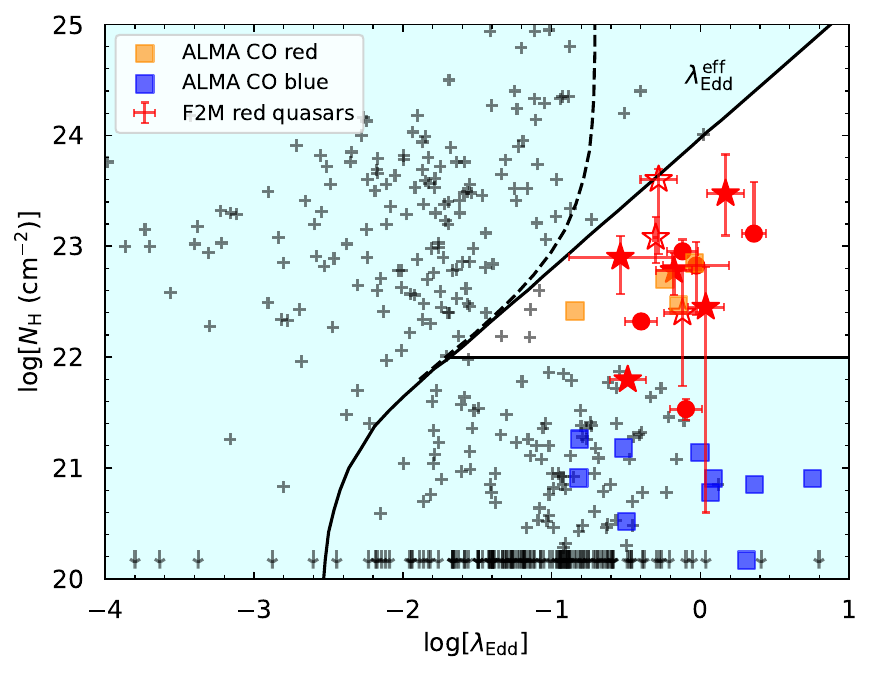}
\includegraphics[scale=0.58]{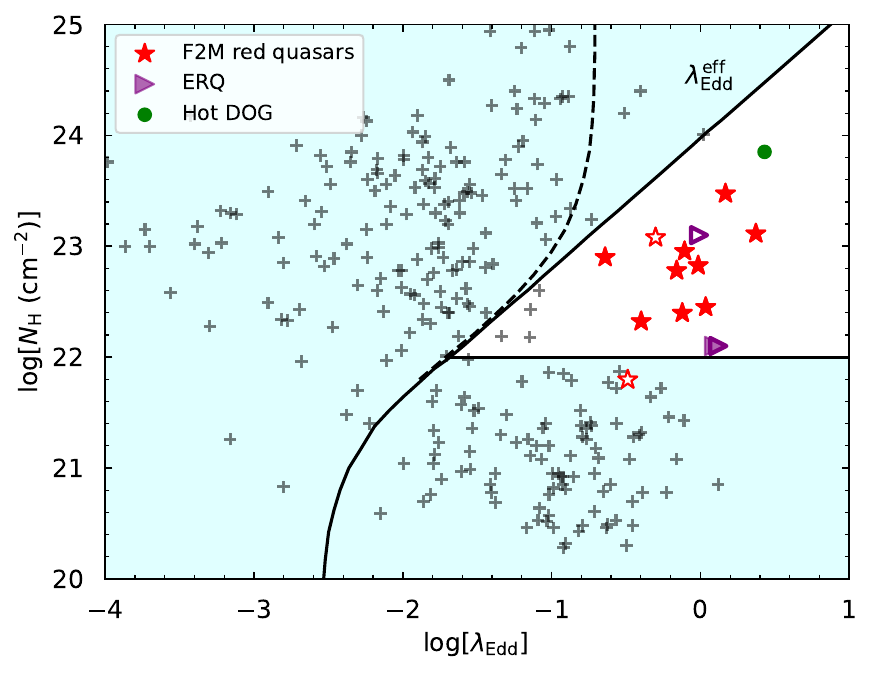}
\caption{Column density, $N_{H}$, vs. Eddington ratio, $\lambda_{Edd}$ for F2M red quasars (red points).
The curved solid line (labelled $\lambda^{\rm eff}_{\rm Edd}$) represents the region where radiation pressure is insufficient to expel the obscuring gas, under the assumption of single scattering, resulting in a high covering fraction \citep{Fabian08}. 
In the white triangular region, the radiation pressure is sufficiently high to blow out the gas and produce outflows. 
The dashed line amends this region by including radiation trapping \citep{Ishibashi18}.
AGNs below $\log(N_{H})$ = 22 may be obscured by dust lanes in their hosts. 
{\em Swift}/BAT AGN in the local universe are shown for comparison \citep[black cross symbols;][]{Ricci17} demonstrating the paucity of sources in the blowout region among normal AGNs. 
{\em Left -- } Filled circles are the four previously analyzed sources from \citetalias{LaMassa16b} and \citetalias{Glikman17a} and F2M~J0915. Filled stars are the five sources in this work whose $N_H$ was determined by spectral modeling. Open stars are the three sources whose $N_H$ was estimated from their HRs. We also plot, for comparison, a sample of 14 quasars studied in CO with ALMA \citep{Stacey22} who found that only the reddened quasars ($E(B-V)>0.5$; orange squares) and none of the unreddened quasars ($E(B-V)<0.1$; blue squares) met blowout conditions. 
{\em Right --} This panel focuses on the F2M red quasars that have host morphologies from high resolution imaging where filled sources are mergers while open sources are undisturbed. 
We also plot, for comparison, three ERQs (purple triangles) that also posses HST imaging and morphological information. The green circle is the Hot DOG from \citet{Ricci17b}. 
While all the objects with merging hosts reside in the blowout region, the region is not exclusively populated by mergers.}
\label{fig:blowout}
\end{figure*}

The key to blowing out gas from the vicinity of an AGN – and possibly out of the host galaxy entirely – may be radiation pressure from a high-enough luminosity pushing against infalling material whose composition is a mixture of partially ionized dust and gas. 
Such a medium has an ‘effective cross section’, $\sigma_i$, that is larger than the Thomson cross section, $\sigma_T$ , reducing the Eddington luminosity for the system. 
This resultant ‘effective Eddington limit’ is thus lower than that for pure ionized hydrogen, enabling AGN with a sufficiently high accretion rate, and a not-too-high column density to blow out the dust and gas \citep{Fabian06,Fabian08,Ishibashi18}.
According to this theory, the interplay between an AGN’s accretion, obscuration, and radiative feedback can be understood through two parameters: $\lambda_{\rm Edd}$ and $N_H$, shown in Figure \ref{fig:blowout}. 
Sources with the right combination of $\lambda_{\rm Edd}$ and $N_H$ are found in the white triangular region, referred to as the ``forbidden'' or ``blow-out'` region, where the luminosity is high enough to produce outflows and gas density is low enough to avoid stalling them.
Since the blow-out of gas as a result of radiation pressure can involve multiple scatterings, as the opacity of the material increases the blowout region can be expanded out to the dashed line, which takes radiation trapping into account \citep{Ishibashi18}. 

\citet{Ricci17} explored the nature of obscuration for {\em Swift}/BAT AGN, which are hard X-ray-selected AGN with $z<0.05$, in the $N_H$ vs. $\lambda_{\rm Edd}$ plane and find that they lie in the region suggestive of long-lasting obscuration whether by dust lanes in the host galaxy for sources with $\log{N_H} < 22$ or by a high covering fraction of nuclear obscuration for sources with $\log{N_H} > 22$. 
These sources are plotted with black crosses in Figure \ref{fig:blowout} confirming that the vast majority of low-luminosity, local AGN are not engaged in radiative feedback. 

The F2M red quasars from \citetalias{LaMassa16b} and \citetalias{Glikman17a} as well as F2M~J0915 are shown with filled red circles. 
The red stars are sources from this work where filled stars are the five high-count objects whose $N_H$ values were determined from spectral fitting and the open stars are the three low-count sources whose $N_H$ values were estimated from HRs. 
The two sources with upper limits on their X-ray counts are not shown. 
One of these low-count sources, F2M~J1106, is on the edge of the region modified by radiation trapping. 
This source shows bi-conal outflowing superbubbles in [\ion{O}{iii}] consistent with trapped photons pushing against an entrained shell of gas expanding out \citep{Shen23}. 
The time-scale for these bubbles is estimated to be $\sim10$ Myrs, which is roughly consistent with the timeline for the most heavily absorbed simulations in \citet{Ishibashi18}.
However, it is also possible that because its $N_H$ value was determined by its HR, this estimate sufficiently uncertain that it may actually live in the unmodified blowout region. 
We note that F2M~J0830, whose X-ray analysis was performed in \citetalias{LaMassa16b}, also shows bi-conal outflowing superbubbles \citep{Shen23}.

All but one (F2M~J1151) of the F2M red quasars in the blowout region show evidence for merging morphologies in their hosts. 
The two source below the $\log(N_{H}) = 22$ line, where the relatively low obscuration is due to dust lanes in the host galaxy, either have undisturbed morphologies (F2M~J1715) or lack imaging (F2M~J1227). 

In an independent investigation of quasar outflow properties at sub-mm wavelengths, \citet{Stacey22} found similar differences between blue and red quasars in archival ALMA observations of sixteen Type 1 quasars with $J \ge 7$ CO lines. 
Four of these sources have $E(B-V)>0.5$ determined from SED fitting, while the remaining sources have $E(B-V)<0.1$. 
We plot the red and blue quasars from \citet{Stacey22} in the left panel of Figure \ref{fig:blowout} with orange and blue squares, respectively. 
Here, too, red quasars with molecular outflows detected by ALMA are in the blowout region. 
Blue quasars without outflows do not reside in blowout region.
In addition, the analysis of the CO lines by \citet{Stacey22} reveals molecular outflows with velocities of $500-1000$ km s$^{-1}$ in the red quasars, while the blue quasars have weaker velocities $\lesssim 300$ km s$^{-1}$. 

Intriguingly, all of the other dust-obscured quasar samples discussed above (HRQs, Hot DOGs, ERQs, red quasars from Stripe 82, as well as the WISSH quasars) that have $N_H$ and $\lambda_{\rm Edd}$ in the literature reside in the blowout region. 
Figure 7 of \citet{Lansbury20} shows the aforementioned sources, including the F2M red quasars from \citetalias{LaMassa16b} and \citetalias{Glikman17a}, on the $N_H$ vs. $\lambda_{\rm Edd}$ diagram.
However, high resolution imaging is limited to only a handful of these sources. 
In the right hand panel of Figure \ref{fig:blowout} we plot $N_H$ vs. $\lambda_{\rm Edd}$ again, focusing only on sources with known host morphologies and we distinguish between merging and undisturbed systems with closed and open symbols, respectively. 
Twelve F2M red quasars are shown with stars where the two that lack {\em HST} imaging, F2M~J1227 and F2M~J1106\footnote{F2M~J1106 does possess GMOS IFU imaging with 0\farcs4 spatial resolution, which is sufficient to reveal galaxy-scale ($>10$ kpc) bubbles with a bi-conal morphology. However, it is unknown whether the quasar is hosted by a merger.}, are omitted. 
While F2M~J1715, which appears to be undisturbed\footnote{We note that the among the 11 F2M quasars with sufficient X-ray counts and imaging to be plotted in the right panel of Figure \ref{fig:blowout}, nine come from the HST observations in  \citet{Urrutia08} that were designed to reach similar depths; the exposure times ranged from $\sim 1600$ s to $\sim 2300$ s in the F814W filters, depending on the redshift. However, F2M~J1715 was observed in an 800 s snapshot observation, which is $\sim 2\times$ shorter than the exposure time for, e.g.,  F2M~J1532, which is at a similar redshift. This means that the surface brightness limit for the image of F2M~J1715 is $\sim 0.8$ mag shallower and some merger features might be missed.}, lies outside the blowout region where dust lanes can explain its properties, F2M~J1151, which is also undisturbed, lies in the blow out region. 

Three ERQs have both HR-based $N_H$ estimates \citep{Goulding18} and $\lambda_{\rm Edd}$ \citep{Perrotta19} as well as HST imaging \citep[J0832+1615, J0834+0159, J1652+1728;][]{Zakamska19}. 
All three reside in the blowout region, shown with purple triangles (two points overlap), but only J1652+1728 is a major merger. 
Though only a few Hot DOGs have X-ray data that enable a measurement of $N_H$ and most are Type 2, precluding a measurement of a black hole mass and thus $\lambda_{\rm Edd}$, we are able to plot one source, WISE~J1036+0449 from \citet{Ricci17b}, in the blowout region with a green circle. Although this object does not have morphological information, as noted above, the overall population of Hot DOGs has a high merger fraction. 

While all sources with merging hosts reside in the blowout region, a merging morphology is not a necessary condition to meet the requirements needed to blow out large amounts of dust and gas. 
The presence of winds and outflows may be a more predictive indicator. 
The entire ERQ sample is shown to have strong outflows in [\ion{O}{iii}], with the highest velocities ($\sim 2000 - 7000$ km s$^{-1}$) in the reddest sources \citep{Perrotta19} and a fourth ERQ source that lacks imaging (J0006+1215) also meets blowout conditions in $N_H$ vs. $\lambda_{\rm Edd}$.
Likewise, the HRQs lack host morphology information but reside in the blowout region \citep[see Figure 7;][]{Lansbury20} and exhibit strong outflows with velocities up to 2500 km s$^{-1}$ in [\ion{O}{iii}] \citep{Temple19}.
In a reverse approach, \citet{Kakkad16} selected AGN in the COSMOS survey that were located in the blowout region and conducted follow-up IFU observations in [\ion{O}{iii}] and find outflow velocities of $\sim 600 - 1900$ km s$^{-1}$.
None the less, a systematic imaging campaign on the various samples of obscured AGN in the blow out region is needed to more thoroughly investigate any connection between mergers and outflows and better understand the conditions under which radiative feedback dominates. 

\section{conclusions} \label{sec:conclusions}
In this paper, we investigate the accretion and obscuration properties of a sample of merger-dominated luminous red quasars via X-ray observations.
This sample consists of 10 newly analysed X-ray observations as well as 5 previously published sources. 
All but two have high resolution imaging with {\em HST} and one of those two has high resolution, high quality IFU imaging in [\ion{O}{iii}].
Although the sources were not chosen to reside in mergers, ten sources have clear evidence of morphologically disturbed hosts (as previously determined by \citealt{Urrutia08} and \citealt{Glikman15}).  
The sample consists of eight new observations, two sources with archival data sets, and five previously published sources. 
When sufficient counts enabled it, we performed spectral modeling to extract parameters such as $N_H$ that enabled a calculation of the absorption-corrected luminosity. 
Lower count objects were analyzed via their HRs to estimate $N_H$; in cases of non-detection, we determine upper limits.  
We combine these X-ray-derived properties with host galaxy morphological information from high resolution imaging, dust reddening, infrared luminosity, and accretion rate ($\lambda_{\rm Edd}$). 
These data allow us to investigate the connection among these properties and we find:
\begin{enumerate}
    \item F2M red quasars have dust-to-gas ratios that are in general lower than the interstellar medium of the Milky Way. Their dust-to-gas ratios are consistent with low-luminosity AGN in the local universe, though the ratio likely arises from very different physics. The dust-to-gas ratios of F2M red quasars is somewhat lower than, but roughly consistent with the dust-to-gas ratios of comparison samples of luminous dusty quasars. 
    \item F2M red quasars are under-luminous in X-rays at a given infrared luminosity when compared with local, low luminosity relations as well as luminous, unreddened sources that straddle the high luminosity relation. However, their X-ray deficit is consistent with other, more luminous, dust-obscured quasars such as the Hot DOGs.
    \item with the exception of two sources, F2M red quasars reside in the ``forbidden'' region of $N_H$ vs. $\lambda_{\rm Edd}$ indicative of them being in a blowout phase due to radiation pressure on dust. Furthermore, all F2M red quasars with merging hosts are in the blowout region as are other luminous dusty quasars from comparison samples. A broader investigation of the host morphologies of blue quasars outside the blowout region is needed to better understand any connection among reddening, feedback, and mergers.
\end{enumerate}

These findings lend further support to F2M red quasars, along with other luminous dust-reddened quasars, being in a brief transitional phase in a merger-driven co-evolution of galaxies and their supermassive black holes. 

\section*{Acknowledgements}

E.G. acknowledges the generous support of the Cottrell Scholar Award through the Research Corporation for Science Advancement. 
E.G. is grateful to the Mittelman Family Foundation for their generous support. 
E.P. and L.Z. acknowledge financial support from the Bando Ricerca Fondamentale INAF 2022 Large Grant ``Toward an holistic view of the Titans: multi-band observations of z>6 QSOs powered by greedy supermassive black-holes.''
We thank Laura Blecha for useful discussions on the nature of F2M~J1507.
We thank Hannah Stacey for providing the data for the ALMA-detected quasars plotted in Figure \ref{fig:blowout}.
We acknowledge the efforts of Charlotte Moore in the initial phase of this project. 

Support for this work was provided by the National Aeronautics and Space Administration through Chandra Award Number 21700216 issued by the Chandra X-ray Center, which is operated by the Smithsonian Astrophysical Observatory for and on behalf of the National Aeronautics Space Administration under contract NAS8-03060.
The scientific results reported in this article are based on observations made by the Chandra X-ray Observatory, data obtained from the Chandra Data Archive, and observations made by the Chandra X-ray Observatory and published previously in cited articles.
This research has made use of software provided by the Chandra X-ray Center (CXC) in the application packages CIAO.
We gratefully acknowledge the National Science Foundation's support of the Keck Northeast Astronomy Consortium's REU program through grant AST-1950797. 

\section*{Data Availability}

The X-ray data underlying this article are publicly available through the {\em Chandra} archives. 




\bibliographystyle{mnras}
\bibliography{ChandraF2M.bbl}




\appendix 
\section{Notes on the Spectral fitting for individual objects} \label{apx:xray_fitting}

\subsection{F2M~1507+3129}\label{appendix:f2m1507}

The Balmer lines of this object have a blue-shifted broad emission component in addition to broad lines at the systemic velocity, which is determined by the [\ion{O}{iii}] lines in its optical spectrum. 
We note that the the  blue-shifted broad emission component of H$\beta$ is similar in structure to H$\alpha$, which is shown in Figure \ref{fig:mbh}. 
We attribute this component to an out-flowing wind, rather than accretion disk geometry, due to the lack of a red-shifted component that is typically seen in double-peaked-emitting AGN. 
The H$\alpha$ component is blue-shifted by 91\AA, corresponding to a velocity of $0.014c$, which is slow compared with typical outflow velocities seen in BALs ($\sim 10,000 - 60,000$ km s$^{-1}$). 
We use the line width from the broad component at the systemic velocity, which we assume to represent virialized motion, to compute the black hole mass.

As is seen in Figure \ref{fig:xray_spectra}, the X-ray spectrum exhibits some unusual features including what appears to be a deficit of flux around $4 - 5$ keV. 
There is also apparent soft excess at energies below 1.5 keV that required a model with a leakage/scattering component (Eqn. \ref{eqn:dpl}). 
However, given the complexity of this model and the small number of spectral bins, we chose to freeze the photon index to $\Gamma = 1.8$. We initially ignored the $4 - 5$ keV and found an acceptable fit with $N_H = 3\times10^{23}$ cm$^{-2}$ and a scattering fraction of 16\%. 
We further interpret the flux deficit at $4 - 5$ keV as absorption of \ion{Fe}{xxvi} Ly$\alpha$ ($E_{\rm rest} = 6.966$ keV) due to an ultra-fast outflow (UFO) and amend the model in Eqn. \ref{eqn:dpl} by adding a Gaussian absorption component. 

\begin{equation}
    {\tt phabs*(zpowerlw + zphabs*zpowerlw + zgauss)}.    \label{eqn:ufo}
\end{equation}

A fit to this model resulted in the same $N_H$ and scattering fraction, while accounting for the absorption at the blue-shifted energy of 4.59 keV, results in a UFO velocity of $0.26c$.
The continuum model components are shown as dotted lines in Figure \ref{fig:xray_spectra}. Table \ref{tab:f2m1507} lists the best-fit parameters for this source.

This UFO velocity is significantly higher than that seen in the blue-shifted Balmer emission and arises from radii closest to the central engine. While these features are therefore not associated with the same outflowing system, their presence may be indicative of feedback on many scales due to sustained outflowing winds and shocks. 
Theoretical models predict that radiation driven relativistic winds interact with shocks against the ISM, triggering galaxy-wide ionized and neutral outflows \citep[e.g.,][]{Zubovas12}.
A higher count X-ray spectrum would allow for a more thorough exploitation of the energy resolution of e.g., XMM-Newton, to better constrain the outflow properties closest to the SMBH. IFU spectroscopy of the Balmer lines would similarly trace the kinematics of the large scale outflows, to fully trace the feedback energy being injected into host galaxy by this quasar. 

\begin{table}
\caption{UFO Model Fit Parameters}
\label{tab:f2m1507}
\begin{tabular}{lr}
\hline
{Parameter} & {Value} \\
\hline
$\Gamma$                  & 1.8{$^\dagger$} \\
Power-law normalization   & $1.1_{-0.2}+^{1.2} \times 10^{-4}$ \\
$N_{H}$ (cm$^{-2}$)       & $3.0_{-1.7}^{+3.7}\times10^{23}$ \\
$E_{Fe XXVI}$ Ly$\alpha$ (keV) & 6.966{$^\dagger$} \\
$\sigma_E$ (keV)          & $2.4\times 10^{-4}${$^\ddagger$}\\
EW (keV)                  & $2.3$ \\
$f_{\rm scatt}$ (\%)      & $16\pm13$ \\
$v_{\rm UFO}$             & $0.26c$ \\
\hline
\multicolumn{2}{l}{$^\dagger$ This parameter was frozen. }\\
\multicolumn{2}{l}{$^\ddagger$ Given that this feature was fit to a region represented by only two}\\ 
\multicolumn{2}{l}{spectral bins, this value is highly uncertain with an unconstrained lower}\\
\multicolumn{2}{l}{bound and an upper bound of $\sigma = 1.45$.  }\\
\multicolumn{2}{l}{energy.}\\
\end{tabular}
\end{table}

\subsection{F2M~1532+2415}\label{appendix:f2m1532}

This source shows emission at 4.1 keV which is the redshifted fluorescent Iron K$\alpha$ line ($E_{\rm rest} = 6.4$ keV) suggestive of significant reflected emission often seen atop a strongly suppressed continuum, which is typical in Type 2 quasars. 
F2M~J1532, however, is a Type 1 source, with broad emission lines seen in its spectrum. 
We performed a self-consistent physically motivated joint fit, with both {\em Chandra} observations, to the X-ray spectrum to properly account for line-of-sight attenuation, scattering of photons into the line of sight, and the fluorescent line emission responsible for the Fe K$\alpha$ feature. 
We employed Equation \ref{eqn:mytorus} in XSpec in so-called `coupled' mode, where the column densities ($N_{H,Z}, N_{H,S}$), torus inclination angle ($\theta_{\rm obs}$), normalizations of the scattering and fluorescent line coefficient components ($A_S$ and $A_L$, respectively) are tied together to preserve the model's self-consistency. This fit, however, yielded poor results in a best-fit inclination angle of $60^\circ$, which is the default, fixed opening angle of the torus-shaped absorber in the MYTorus model. This suggests a grazing incidence angle which is highly unlikely. Therefore, while this is a statistically acceptable fit, it is not a physically meaningful result \citep[see discussion in][for more details on this phenomenon]{LaMassa16a,LaMassa23}. 

Under such circumstances, it is advisable to fit the spectrum with the same MYTorus model (Eqn. \ref{eqn:mytorus}) but in `decoupled' mode.  This approach assumes that the absorbing and X-ray reprocessing media are not necessarily the same, nor are they smooth and homogeneous, as is assumed with a simple torus model. In this approach the line-of-sight column density ($N_{\rm H,los}$) is provided by the $N_{H,Z}$ parameter which is decoupled from the global column density ($N_{\rm H,global}$) from surrounding clouds, provided by the  $N_{H,S}$ terms in Equation \ref{eqn:mytorus}, which are still tied to each other. The inclination angles are also decoupled, such that the transmitted component is frozen at $90^\circ$, while the scattered and fluorescent line components are frozen to $0^\circ$. In this model, we are not assuming a homogeneous donut-shaped absorber, but utilize the radiative transfer determined by MYTorus to consider light passing through and scattering off of a clumpy and patchy medium surrounding the AGN. 

A fit to this model yields reassuring results. The power-law is best described by $\Gamma = 1.4$ which, while at the flat end of the range of indices for AGN, is consistent with the value found from the phenomenological XSpec model (Eqn. \ref{eqn:dpl}). Additionally, the best-fit line-of-sight column density is $N_{\rm H,Z} = 7.4\times 10^{22}$ cm$^{-2}$, which is also consistent with the value found from the phenomenological XSpec model (Eqn. \ref{eqn:dpl}; $N_H = 7.9\times 10^{22}$ cm$^{-2}$) which only considers line-of-sight absorption and does not account for additional physics. The scattering fraction is also consistent, with the MYTorus fit yielding $f_{scatt} = 10\%$ compared with 11\% in the phenomenological model. Finally, the best-fit global column density  is $N_{\rm H,global} = 10^{24}$ cm$^{-2}$, in the Compton thick regime, which means that this scattered component does not significantly contribute to the continuum allowing for the strong similarity seen with the phenomenological model while also accounting for the Fe K$\alpha$ line. This suggests that F2M~J1532 is enshrouded by a heavy amount of absorption which is non uniform and our line-of-sight happens to coincide with an opening allowing a direct view to the broad line emission from the quasar.

\begin{figure*}
    \centering
    \includegraphics[scale=0.31]{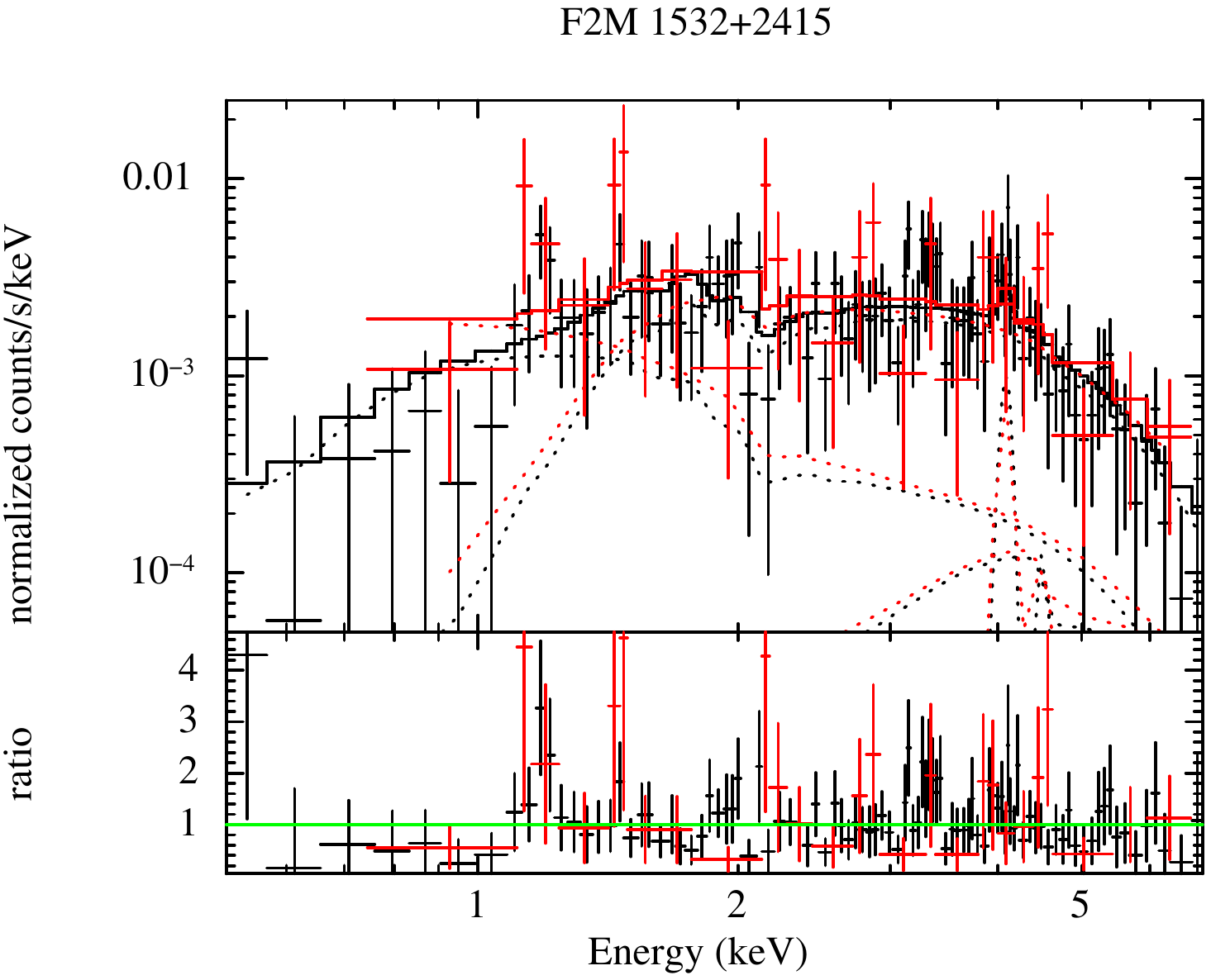}
    \includegraphics[scale=0.31]{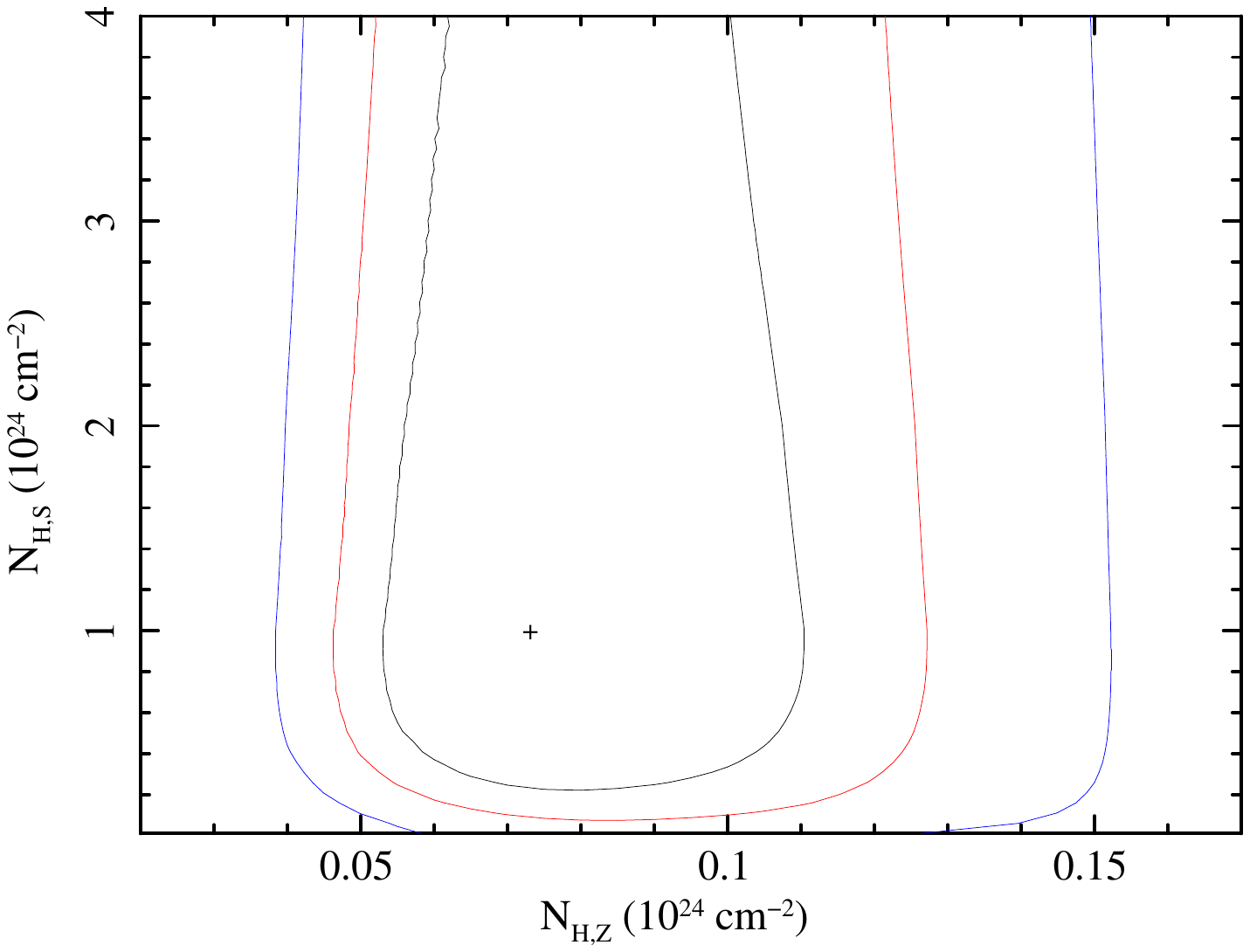}
    \caption{Left -- MYTorus plus scattered power-law (Eqn \ref{eqn:mytorus}) in a `decoupled' joint fit to the observations of F2M~J1532.  Black points and lines represent ObsID 3138 and red points and lines represent ObsID 3338. The solid line shows the combined model, while the dotted lines are the individual model components. 
    Right -- $\chi^2$ contour plot of the global column density ($N_{H,S}$) vs. the line-of-sight column density ($N_{H,Z}$) from the `decoupled' MYTorus fit. The cross represents the best-fit parameters at $N_{H,Z} = 7.4\times10^{22}$ cm$^{-2}$ and $N_{H,S} = 1.0\times10^{24}$ cm$^{-2}$.  The black, red, and blue curves show the 68\%, 90\%, and 99\% confidence levels. We see that while $N_{H,Z}$ is reasonably well-constrained and consistent with the value found for the phenomenological model (Eqn. \ref{eqn:dpl}) of $N_{H,Z} = 7.9\times10^{22}$ cm$^{-2}$, $N_{H,S}$ is poorly constrained. }
    \label{fig:mytorus_f2m1532}
\end{figure*}

Table \ref{tab:f2m1532} lists the best-fit MYTorus parameters for this source. Figure \ref{fig:mytorus_f2m1532} shows the best fit MYTorus model plotted atop the source spectrum, with the individual model components shown separately on the left and a contour plot of the global ($N_{H,S}$) versus line-of-sight ($N_{H,Z}$) column densities showing that while $N_{H,Z}$ is well constrained and consistent with the phenomenological XSpec model, the global column density is poorly constrained and highly uncertain. 

\begin{table}
\caption{Decoupled MYTorus Fit Parameters}
\label{tab:f2m1532}
\begin{tabular}{lr}
\hline
{Parameter} & Value \\
\hline
$\Gamma${$^\dagger$}                & $1.41_{-0.01}^{+0.28}$ \\
Power-law normalization             & $1.0_{-0.2}^{+1.3} \times 10^{-4}$ \\
$N_{H,Z}$ (cm$^{-2}$) line-of-sight & $7.4_{-2.5}^{+3.7} \times 10^{22}$ \\
$N_{H,S}$ (cm$^{-2}$) global{$^\ddagger$} & $1.0_{-0.8} \times 10^{24}$ \\
$f_{\rm scatt}$ (\%)                & $9.5_{-4.6}^{+3.4}$ \\
C-stat (dof)                        & 106.99 (119) \\
\hline
\multicolumn{2}{l}{$^\dagger$This parameter was constrained with a lower bound of $\Gamma \ge 1.4$.}\\
\multicolumn{2}{l}{$^\dagger$ The error analysis of this parameter was found to have an unconstrained} \\
\multicolumn{2}{l}{upper bound, as illustrated by the contour diagram shown in Figure \ref{fig:mytorus_f2m1532}.}
\end{tabular}
\end{table}


\bsp	
\label{lastpage}
\end{document}